\documentclass[10pt,a4paper]{article}

\usepackage{arxiv}
\usepackage{tikz}
\usepackage[utf8]{inputenc} 
\usepackage[T1]{fontenc}    
\usepackage{hyperref}       
\usepackage{url}            
\usepackage{booktabs}       
\usepackage{amsfonts}       
\usepackage{nicefrac}       
\usepackage{microtype}      
\usepackage{lipsum}
\usepackage{miller}

\usetikzlibrary{matrix, positioning}
\usetikzlibrary{calc}
\usepackage{amssymb}
\usepackage{lineno,hyperref}
\usepackage{amsmath}
\usepackage{bm}
\usepackage{subcaption}
\usepackage{graphicx}
\usepackage{svg}
\usepackage{rotating}
\usepackage{lscape}
\usepackage{array,multirow}
\usepackage[title]{appendix}
\usepackage{float}
\usepackage{stmaryrd}
\usepackage{textcomp}
\usepackage{algorithm}
\usepackage{algpseudocode}
\usepackage{amsthm}
\usepackage{listings}
\usepackage{xcolor}
\usepackage{siunitx}

\usepackage{pifont}
\usepackage{lineno}
\usepackage{comment}
\usepackage{import}
\usepackage{physics}
\usepackage{comment}
\usetikzlibrary{positioning}

    \DeclareFontFamily{OT1}{pzc}{}
\DeclareFontShape{OT1}{pzc}{m}{it}{<-> s * [1.10] pzcmi7t}{}
\DeclareMathAlphabet{\mathpzc}{OT1}{pzc}{m}{it}
\usepackage[numbers]{natbib}
\usepackage[algo2e,ruled,vlined]{algorithm2e}
\DontPrintSemicolon

\newtheorem{remark}{Remark}
\title{Deep Convolutional Ritz Method: Parametric PDE surrogates without labeled data}

\author{Jan N. Fuhg \\
Cornell University\\ Ithaca, NY, USA \\
\And
Arnav Karmarkar \\
Cornell University\\ Ithaca, NY, USA \\
\And
Teeratorn Kadeethum\\
Sandia National Laboratories \\
Albuquerque, NM,  USA \\
\And
Hongkyu Yoon\\
Sandia National Laboratories \\
Albuquerque, NM,  USA \\
\And
Nikolaos Bouklas \\
Cornell University\\ Ithaca, NY, USA \\
}

\DeclareMathOperator*{\argmin}{arg\,min}

\begin{document}
\maketitle

\begin{abstract}
Parametric surrogate models for partial differential equations (PDEs) are a necessary component for many applications in the computational sciences, and convolutional neural networks (CNNs) have proved as an excellent tool to generate these surrogates when parametric fields are present. CNNs are commonly trained on labeled data based on one-to-one sets of parameter-input and PDE-output fields.
Recently, residual-based convolutional physics-informed neural network (CPINN) solvers for parametric PDEs have been proposed to build surrogates without the need for labeled data. These allow for the generation of surrogates without an expensive offline-phase.
In this work, we present an alternative formulation termed Deep Convolutional Ritz Method (DCRM) as a parametric PDE solver. The approach is based on the minimization of energy functionals, which lowers the order of the differential operators compared to residual-based methods. 
Based on studies involving the Poisson equation with a spatially parameterized source term and boundary conditions, we found that CNNs trained on labeled data outperform CPINNs in convergence speed and generalization ability.
Surrogates generated from DCRM, however, converge significantly faster than their CPINN counterparts and prove to generalize faster and better than surrogates obtained from both CNNs trained on labeled data and CPINNs.
This hints that DCRM could make PDE solution surrogates trained without labeled data possible.
\end{abstract}

\section{Introduction}
Partial differential equations (PDEs) are omnipresent in engineering science and play a significant role in describing physical problems.
In this context, many computational tasks involve repeated evaluations of PDEs given statistically similar parameterized inputs.
These parametric solutions to PDEs are of interest in different application areas such as
topology optimization \cite{gogu2015improving,xia2014reduced,keshavarzzadeh2021robust},
uncertainty quantification \cite{roache1997quantification,chen2017reduced,tripathy2018deep},
PDE-constrained optimization problems \cite{biegler2003large,fahl2003reduced,zahr2015progressive}
and
multiscale problems \cite{fuhg2021modeldatadriven,fuhg2022local, fuhg2022physics}.
Due to the excessive computational costs of high-fidelity numerical solvers such as the Finite-Element \citep{wriggers2008nonlinear} or Finite-Volume method \citep{moukalled2016finite}, surrogate modeling (also known as metamodeling or reduced order modeling (ROM)) approaches have been applied to obtain a quick-to-evaluate model that (ideally) allows one to obtain a prediction of the PDE solution (given an input) in a fraction of the time that high-fidelity numerical solvers take.
Prominently, these surrogates are based on the offline-online paradigm,
a methodology using labeled data, where in the offline phase datasets are generated from high-fidelity simulations that are utilized for training the model and building the basis for the online prediction phase.
Furthermore, intrusive and non-intrusive models can be distinguished.
Intrusive models typically use projections of the full order model (i.e., from the high-fidelity solver) onto a reduced space where access to the discretized PDE operators is required. In this context, the most popular intrusive ROM method uses Proper Orthogonal Decomposition (POD) with a Galerkin projection (POD-Galerkin) \citep{berkooz1993proper,couplet2005calibrated}.
ROMs that do not rely on discretized PDE operators are referred to as non-intrusive models.
These models directly learn the input-output mapping from the high-fidelity simulation data.
Different machine learning approaches have been applied to generate non-intrusive surrogates for parametric PDEs such as Gaussian process regression \citep{guo2018reduced,ortali2020gaussian}, feedforward neural networks (FNNs) \citep{tripathy2018deep,bhattacharya2020model}, convolutional neural networks (CNNs) \citep{zhu2018bayesian,kutyniok2022theoretical,khoo2021solving}, generative adversarial networks \cite{sun2018discovering,kadeethum2021framework} and pure data-driven operator-based techniques \citep{li2020neural,li2020fourier}.
Since the numerical models are often computationally expensive, adaptive (or active) sampling methods have been developed that iteratively aim to find the set of input values which, when evaluated and concatenated to the training dataset, best
enhance the accuracy of the surrogate model  \cite{liu2018survey,fuhg2021state}. These are commonly applied to scalar-valued parameter inputs \cite{fuhg2022classification} as well as input fields for which known parametrizations exist \cite{schobi2015polynomial}.

In the last few years, physics-informed machine learning (PIML) techniques have had a significant impact on computational science and engineering. In PIML, the training space of machine learning models is constrained by the physical knowledge of the underlying system and more specifically the PDEs at hand.
This can mainly take two forms: (i) Traditional data-driven reduced order models are enhanced by combining the information of the labeled data (e.g., loss between model prediction and ground truth PDE solution) with physical constraints (e.g., enforcement of conservation laws, fulfillment of boundary conditions) \cite{wang2020towards,mohan2020embedding}. (ii) Machine learning techniques such as neural networks are used as full PDE-solvers using automatic differentiation \citep{griewank1989automatic} and can be used with or without labeled data.
A realization of the latter is Physics-Informed Neural Networks (PINNs) which use point-wise residual information of the PDE to tune the trainable parameters of a feed-forward neural network such that the outputs of the network at independent parameter inputs approximate the solution of an initial-boundary value problem \cite{lagaris1998artificial,raissi2019physics,wessels2020neural,fuhg2022interval}. These were initially developed for forward and inverse problems involving non-parametric PDEs; namely, the forward solver could only obtain the solution for a fixed spatial distribution of PDE coefficients (which could for example correspond to spatially varying material parameters, body forces) similar to what a Finite Element solver would be able to achieve. The distinguishing feature of PINNs was the ability to incorporate limited labeled data towards the solution of inverse problems.
Later, different machine learning-based techniques were developed to use PDE residuals as means towards building surrogates for parametric PDEs.
In this context, DeepONets with labeled data (e.g., from numerical simulations)
 \cite{lu2019deeponet,lu2021learning} and without labeled data \citep{wang2021learning} can be seen as an extension to the PINN architecture.
Parametric PDEs have also been solved using CNN-parametrizations of input fields. In that setting, the PDE residual is obtained by using a finite-difference layer at the output of the network, avoiding automatic differentiation for the approximation of the differential operators of the PDE. Since residual information is necessary to constrain the trainable parameters, we term these models Convolutional Physics-Informed Neural Networks (CPINNs).
CPINNs have successfully been deployed for stochastic PDEs
\citep{zhu2019physics}, irregular domains \citep{gao2021phygeonet} and as spatiotemporal forward-solvers \citep{ren2022phycrnet}. The advantage of these approaches is their ease of implementation and input field parametrization compared to other existing approaches. 

In contrast to these residual-based approaches, PDE solvers based on energy minimization principles have been studied. Using feedforward neural networks The Deep Ritz Method (DRM) \citep{yu2018deep} and extensions  \cite{yu2018deep,liao2019deep,duan2021convergence} are used for the numerical solution of variational problems.
A similar approach termed
Deep Energy Method \cite{samaniego2020energy, fuhg2022mixed} has been employed in the field of computational mechanics.
Energy-based approaches have a significant advantage in comparison to PDE residual architectures because they require lower order differential operators and therefore show faster convergence behavior and have fewer issues  with vanishing or exploding gradients than PINNs \cite{krishnapriyan2021characterizing,wang2022and}.
However, none of the presented energy-based approaches are designed for generating surrogates for parametric PDEs but instead act as simple forward solvers.

Hence, in this work, we develop an energy-based solver for parametric PDEs using CNNs. We term this approach Deep Convolutional Ritz Method (DCRM). The proposed technique is studied on the two-dimensional Poisson equation with a spatially parameterized source term and boundary conditions for proof of concept.
We compare the method to surrogate models obtained from pure CNNs with labeled data and CPINNs without labeled data and find that the generalization error of DCRM models converges significantly faster than both CNNs and CPINNs and shows improved generalization capabilities on unseen input data.

The paper is structured as follows.
Section \ref{sec::2} provides an overview of surrogate modeling of parametric PDEs, including residual and energy formulations. Furthermore, the basic concepts of CNNs (for reduced order modeling of PDEs) and CPINNs are summarized and explained using the Poisson equation as an example.
In section \ref{sec::3}, we propose  the Deep Convolutional Ritz Method and explain all its components.
The treatment of boundary conditions for CNN-like structures in the context of PDEs is explained in section \ref{sec::4}.
The method is applied to modeling the parameterized Poisson equation and compared to CNN and CPINN surrogate modeling of PDEs in
section \ref{sec::5}.
Lastly, section \ref{sec::6} discusses the results and concludes the work.

\section{Surrogate modeling for parametric PDEs}\label{sec::2}

Consider a time-independent PDE given by
\begin{equation}\label{eq::GeneralPDE}
\begin{aligned}
        \mathcal{N}(u(\bm{x}), \ldots) &= 0,  &&\bm{x} \in \Omega
\end{aligned}
\end{equation}
defined by a differential operator $\mathcal{N}$. Here, $u$ must satisfy both Dirichlet and Neumann boundary conditions.
This is known as the strong form of the PDE and is the functional form that is typically utilized by PINN formulations. \\
If the Euler-Lagrange equation of an energy functional $\mathcal{E}$ coincides with the PDE of eq. (\ref{eq::GeneralPDE}) (c.f. \cite{anderson1980existence}), the solution of the PDEs can alternatively be found by utilizing the minimization of an energy
\begin{equation}\label{eq::GeneralEnergy}
    \min_{u} E(u(\bm{x}), \ldots) = \int \mathcal{E} (u(\bm{x}) , \ldots).
\end{equation}
Here, $u$ is constrained by the same set of boundary conditions as eq. (\ref{eq::GeneralPDE}).
In the context of using numerical techniques (such as neural networks) to find solutions to PDEs, the direct minimization of the energy function of eq. (\ref{eq::GeneralEnergy}) has a significant advantage in comparison to finding solutions based on the strong form (eq. (\ref{eq::GeneralPDE})) since lower order differential operators are needed (since obtaining the strong form from eq. (\ref{eq::GeneralEnergy}) involves differentiation of the functional $\mathcal{I}$)

\begin{remark}
Variational forms of the energy  eq. (\ref{eq::GeneralEnergy}) are the starting point for numerical techniques like the Finite-Element method \cite{reddy2017energy} and Virtual Element Method \cite{beirao2014hitchhiker}. For systems that are based on conservation laws - e.g., energy, mass, momentum - the energy functional can typically be derived from physical understanding. Alternatively, the problem of finding an energy functional
is known as the \textit{"inverse problem of the calculus of variations"} and has received a lot of attention over the years \citep{douglas1941solution,takens1979global,zenkov2015inverse}.
For an overview of different energy functionals, the calculus of variations, and some physical examples, we refer to \cite{weinstock1974calculus}. 
It needs to be pointed out that not all PDEs admit an energy functional.
\end{remark}
In this work we focus on the Poisson equation
\begin{equation}\label{eq::Poisson}
\begin{aligned}
      - \Delta u(\bm{x}) &= f(\bm{x}), \qquad &&\text{in}\, \Omega =[0,1]^{2} \\
       u &= g_{D}, \qquad &&\text{on}\, \partial\Omega_{D} \\
              \frac{\partial u}{\partial \bm{n}} &= g_{N}, \qquad &&\text{on}\, \partial\Omega_{N} \\
\end{aligned}
\end{equation}
where we assume $\partial\Omega_{D} \cup \partial\Omega_{N} = \partial\Omega $, where $\partial\Omega_{D}$  refers to the part of the boundary where Dirichlet boundary conditions are applied  and $\partial\Omega_{N}$  where Neumann boundary conditions are  applied; we note that these two regions are distinct. We are interested in obtaining a deterministic surrogate model for eq. (\ref{eq::Poisson}) for a parameterized source term $f(\bm{x})$ and parameterized boundary conditions $g_{D}$, where both can vary spatially.
It can be shown (see \cite{evans1998partial}) that the energy functional that yields the same solution as eq. (\ref{eq::Poisson}) is given by
\begin{equation}\label{eq::EnergyPoisson}
    E[u] = \int_{\Omega} \frac{1}{2} \norm{\nabla u}^{2} - \int_{\Omega} u f  - \int_{\partial \Omega_{N}} u g_{N}  .
\end{equation}
As highlighted before the variational energy formulation involves only first order differential operators, one order lower that the strong form (eq. (\ref{eq::Poisson})). \\

A pure data-driven approach would utilize the following dataset consisting of $N$ samples to generate a surrogate model for this application
\begin{equation}\label{eq::PoissonData}
    \mathcal{D} = \lbrace \underbrace{\left( \bm{F}_{i}, \bm{G}^{1}_{i}, \bm{G}^{2}_{i}, \ldots \right)}_{\text{inputs}}, \underbrace{\bm{U}_{i}}_{\text{output}} \rbrace_{i=1}^{N}
\end{equation}
where $\bm{F}$ is a source term field, $\bm{G}^{k}_{i}$ is the $k$-th descriptor of the boundary condition of the $i$-th sample and $\bm{U}$ is the output field.
The set of the source term and boundary condition descriptors define the input to the reduced order model while the output is given by the primary output field $\bm{U}$.

When working with CNNs (c.f. Section \ref{sec::CNNs}) the dataset has to be converted into an image-like format. Standard numerical PDE solvers already provide and necessitate discretized versions of the source and output fields which (if necessary) can be interpolated into uniform grids of height $\mathbb{DOF_X}$ and width $\mathbb{DOF_Y}$. 
Parametrization of the boundary condition descriptors ($\bm{G}^{1}_{i}, \bm{G}^{2}_{i}, \ldots$) into images that can be processed by a CNN to learn the spatial correlation of changing boundary conditions is more challenging.
One possible option which can also provide spatial context information is masking functions \citep{alguacil2021effects} of the form
\begin{equation}
    \mathbb{I}_{g}(\bm{x}) = \begin{cases}
    g(\bm{x}), & \text{if } \bm{x} \in \partial \Omega \\
    0.0, & \text{otherwise}
    \end{cases}
\end{equation}
which generate images that project the boundary condition onto its correct spatial position at the border of the image while zeroing all pixels that are not part of the boundary.
Distinctions between types of boundary conditions (Neumann or Dirichlet) can be made using an additional masking function
\begin{equation}
    \mathbb{I}_{t}(\bm{x}) = \begin{cases}
    1.0, & \text{if \,\, } \bm{x} \in \partial \Omega_D \\
    2.0, & \text{if\,\,} \bm{x} \in \partial \Omega_N 
    \end{cases}
\end{equation}
which indicates the spatial positions of the boundary condition types at the image borders. Other common ways of imposing boundary conditions in CNN-based models are discussed in \cite{alguacil2021effects}.
Without loss of generality, we limit ourselves in the following to Dirichlet boundary conditions which can be described using a single masked boundary image.
Hence, in this case given labeled data from PDE solvers the input images can be structured into a 4D tensor of dimension {[}$N$,  $\mathbb{C}_{in}=2$, $\mathbb{DOF_X}$, $\mathbb{DOF_Y}${]}. Similarly, the 4D output tensor is of size {[}$N$, $\mathbb{C}_{out}=1$, $\mathbb{DOF_X}$, $\mathbb{DOF_Y}${]}. Here, $\mathbb{C}_{in}$ and $\mathbb{C}_{out}$ respectively denote the number of input and output channels.

We can define the training dataset
\begin{equation}
    \mathcal{D}_{I} = \lbrace \underbrace{\left( \bm{F}_{i}, \bm{G}_{i} \right)}_{\text{inputs}}, \underbrace{\bm{U}_{i}}_{\text{output}} \rbrace_{i=1}^{N}
\end{equation}
where $\bm{F},\bm{G},\bm{U} \in \mathbb{R}^{N \times 1 \times \mathbb{DOF_X}\times \mathbb{DOF_Y}}$.
A CNN can now be trained to obtain a surrogate model for the parameterized Poisson equation. In cases where no labeled data (i.e., only the input data) is available finite-difference (FD) stencils can be appended to the CNN to discretize and solve the PDE and therefore learn the structure of the PDE.
In order to simplify notation consider $\mathbb{DOF} = \mathbb{DOF_X} = \mathbb{DOF_Y}$ in the following.

Recently, \citep{zhu2019physics} and \citep{gao2021phygeonet} used CNNs with FD on the strong form the PDE eqs. (\ref{eq::GeneralPDE}) and (\ref{eq::Poisson}) to obtain the surrogate (in this work we refer to this approach as CPINN). As an alternative, to CPINN we propose DCRM (c.f. Section \ref{sec::3}) which instead utilizes the energy formulation of eqs. (\ref{eq::GeneralEnergy}) and (\ref{eq::EnergyPoisson}) as a starting point to derive surrogates for parametric PDEs without labeled data.
In the following, we shortly summarize the components of a CNN, define the specific architecture used in this work, and elaborate on the CPINN approach.

\subsection{Data-driven approach using CNNs and labeled data}\label{sec::CNNs}
Convolutional Neural Networks have been widely applied in image-based learning applications due
to their ability to capture long-range spatial correlation and
to efficiently share parameters (especially in comparison to feedforward neural networks). They offer a convenient parametrization of PDE input and output fields
and hence allow one to coherently obtain surrogate models for parametrized PDEs.
One negative aspect of CNNs in the context of solving PDEs is that they (generally) require cartesian uniform grids and rectangular domains as input and output fields. However, efforts have been made to extend CNNs to non-euclidean and non-uniform grids \citep{masci2015geodesic,qi2017pointnet,jiang2019convolutional}. \\

In their basic form, CNNs are multilayer networks built around the convolutional operation \citep{gu2018recent}. The first  and last layers of the network, are input and output layers, respectively.
The network value at the $l$-th layer in the $k$-th channel at the node location $(i,j)$ is obtainable by
\begin{equation}\label{eq::ConvFilter}
    z_{k,i,j}^{l} = \bm{w}_{k}^{l, T} \bm{x}_{i,j}^{l} + b_{k}^{l}
\end{equation}
where $\bm{w}_{k}^{l}$ and $b_{k}^{l}$  are the shared weights (also known as kernel or filter) and biases of the $k$-th channel and $\bm{x}_{i,j}^{l}$ is the input patch around pixel location $(i,j)$.
The trainable parameters can be summarized by the set $\bm{\theta} = \lbrace  \cup_{\forall k, l } (\bm{w}_{k}^{l}, b_{k}^{l}) \rbrace$.

\begin{figure}
    \centering

\begin{tikzpicture}[line join=round,transform shape=0.8]
\def\a{{{6,0,3,2,5},{5,3,1,0,6},{2,0,1,1,1},{4,1,2,1,4},{1,2,1,1,9}}}
\def\b{{{1,4,0},{0,1,2},{1,0,1}}}
\def\c{{{-5,-1,2},{1,1,-9},{-4,8,16}}}

\pgfdeclarelayer{Original layer}
\pgfdeclarelayer{Filter layer}
\pgfdeclarelayer{Result layer}
\pgfsetlayers{Filter layer,main,Original layer,Result layer}

\begin{pgfonlayer}{Original layer}
\begin{scope}[yslant=-.25,local bounding box=Ori]
\fill[yellow, opacity=0.2] (2,2) rectangle +(3,3);
\draw[blue] (0,0) grid (5,5);
\draw[blue,very thick] (0,0) rectangle (5,5);
\foreach \i in {0,...,4}
\foreach \j in {0,...,4}{
\pgfmathsetmacro{\x}{int(\a[\i][\j])}
\path (\i+.5,\j+.5) node{\x};}
\path
(2,2) coordinate (A1)
(2,5) coordinate (A2)
(5,5) coordinate (A3)
(5,2) coordinate (A4);
\path (Ori.north) node[above=2mm]{Original};
\end{scope}
\end{pgfonlayer}

\begin{pgfonlayer}{Filter layer}
\begin{scope}[yslant=-.25,shift={(6,4)},
local bounding box=Fil]
\fill[yellow, opacity=0.2] (0,0) rectangle +(3,3);
\draw[blue] (0,0) grid (3,3);
\draw[blue,very thick] (0,0) rectangle (3,3);
\foreach \i in {0,...,2}
\foreach \j in {0,...,2}{
\pgfmathsetmacro{\x}{int(\b[\i][\j])}
\path (\i+.5,\j+.5) node{\x};}
\path
(0,0) coordinate (B1)
(0,3) coordinate (B2)
(3,3) coordinate (B3)
(3,0) coordinate (B4);
\path (Fil.north) node[above=3mm]{Filter};
\end{scope}
\end{pgfonlayer}

\begin{pgfonlayer}{Result layer}
\begin{scope}[yslant=-.25,shift={(10,6)},
local bounding box=Res]
\fill[yellow, opacity=0.2] (2,2) rectangle +(1,1);
\draw[blue] (0,0) grid (3,3);
\draw[blue,very thick] (0,0) rectangle (3,3);
\foreach \i in {0,...,2}
\foreach \j in {0,...,2}{
\pgfmathsetmacro{\x}{int(\c[\i][\j])}
{\ifthenelse{\i=2 \AND\j=2}{\path (\i+.5,\j+.5) node{\x};}{
\path (\i+.5,\j+.5) node{-};};};}
\path
(2,2) coordinate (C1)
(2,3) coordinate (C2)
(3,3) coordinate (C3)
(3,2) coordinate (C4);
\path (Res.north) node[above=2mm]{Result};
\end{scope}
\end{pgfonlayer}

\foreach \i in {1,2,3,4}
\draw[red,dotted,thick] (A\i)--(B\i);
\foreach \i in {1,2,3,4}
\draw[green,dotted,thick] (B\i)--(C\i);
\end{tikzpicture}
    \caption{Convolution operation with a $3\times 3$ filter and no bias term.}
    \label{fig:convOperation}
\end{figure}
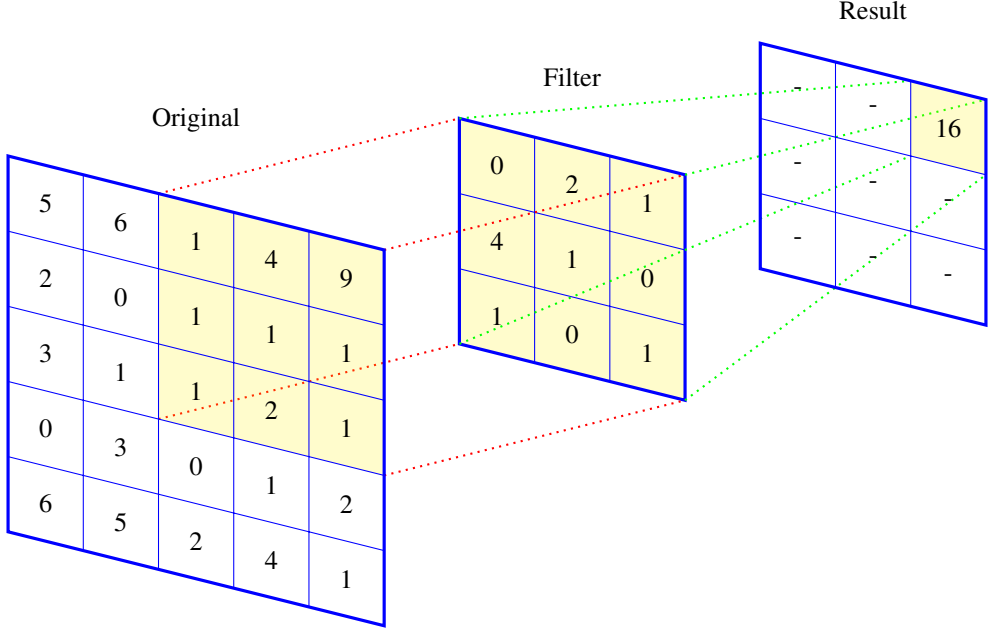

One such operation with a single channel is schematized in Figure \ref{fig:convOperation} where the bias term is not considered.
The mapping can be enhanced by introducing nonlinearities to CNNs by defining an activation function $\sigma(\bullet)$ which takes $z_{k,i,j}^{l}$ as an input and returns the activation value $a_{k,i,j}^{l}$.
\begin{equation}\label{eq::acti}
    a_{k,i,j}^{l} = \sigma (z_{k,i,j}^{l})
\end{equation}
Two typical activation functions are ReLU
\begin{equation}
    a_{k,i,j} = \max(z_{k,i,j}, 0)
\end{equation}
and Leaky ReLU
\begin{equation}
    a_{k,i,j} = \max(z_{k,i,j}, 0) + \lambda \min(z_{k,i,j}, 0)
\end{equation}
where $\lambda \in (0,1)$.

Over the years, the basic CNN architecture of combining eqs. (\ref{eq::ConvFilter}) and (\ref{eq::acti}) has been gradually improved by the addition of other features that increase the robustness of CNNs.
Pooling layers have been added to lower the computational burden of CNN forward and backward propagations and to aid generalization \citep{estrach2014signal}. One commonly applied pooling layer is known as Max-pooling which reads
\begin{equation}
    y_{k,i,j} = \max \left[ \sum_{(m,n) \in \mathcal{R}_{i,j}} a_{k,m,n} \right]
\end{equation}
where $\mathcal{R}_{i,j}$ is a local neighborhood around $(i,j)$. The opposite operation of pooling is known as upsampling.
Other features of CNNs include the dropout operation introduced by \cite{hinton2012improving,srivastava2014dropout} which takes
an input tensor and sets some of its element values to zero with probability $p$ based on Bernoulli distributed samples.
It has been shown that the operation can be used to effectively
reduce overfitting \citep{hinton2012improving}.
Similarly, batch normalization re-centers and re-scales the inputs of each individual layer which has shown to be advantageous, see \cite{ioffe2015batch}.

In this work, we utilize a well-established CNN architecture known as UNet \citep{ronneberger2015u}, which shape we here adopt from \cite{kadeethum2021framework} where it was shown to proficiently work as a data-driven ROM approach for parametrized PDEs. The structure employs an encoder-decoder framework with skip connections \citep{mao2016image} which allows to project the main features linking the inputs and outputs into a latent space similar to autoencoders \cite{wang2014generalized,badrinarayanan2017segnet}. This enables the network to ignore insignificant data (i.e., the zeros of the boundary mask layers) and generalize easier \citep{hinton2006reducing}. Since, in contrast to autoencoders, the input data of the network differs from its outputs, the network can be better described as a heteroencoder \cite{roweis1999linear,bridgman2022heteroencoder}.
Two basic components build this network: (i) A contracting block primarily composed of two convolutional layers followed by a maxpooling operation. The outputs of the convolutional layers are subjected to a Leaky ReLU activation function with $\lambda=0.2$. 
(ii) An expanding block that consists of an upsampling layer and two consecutive convolutional layers, each transformed by a ReLU activation.
The encoder-decoder framework is then defined by an input convolutional layer followed by six encoding, six decoding blocks, and an output convolution.
We choose a kernel size of $3\times 3$ with padding and stride set to $1$.
The architecture of the network is detailed in
Table \ref{tab:unet} which also provides extra information about each layer, such as input and output sizes, activation functions, and where Dropout and Batch normalizations are used. All of the hyperparameters were chosen according to the ones used in \cite{kadeethum2021framework}.

\begin{table}[!ht]
\centering
\caption{Detailed description of the UNet used in this study, input and output sizes are represented by {[}$N$, $\mathbb{C}_{in}$, $\mathbb{DOF_X}$, $\mathbb{DOF_Y}${]} and {[}$N$, $\mathbb{C}_{out}$, $\mathbb{DOF_X}$, $\mathbb{DOF_Y}${]} respectively where $\mathbb{C}_{in}=2$ and $\mathbb{C}_{out}=1$.}
\label{tab:unet}
\vspace{0.5cm}
\begin{tabular}{||l|c|c|c|c|c||}
\hline
Block                  & \multicolumn{1}{c|}{Input size} & \multicolumn{1}{c|}{Output size} & \multicolumn{1}{c|}{Dropout} & \multicolumn{1}{c|}{BN}   & \multicolumn{1}{c|}{Activation}  \\ [0.5ex]  \hline  \hline
$1^{\mathrm{st}}$ convolutional layer & {[}$N$, $\mathbb{C}_{in}$, 128, 128{]}            & {[}$N$, 32, 128, 128{]}            &    -                                     &      -        & Linear                \\ \hline
$1^{\mathrm{st}}$ contracting block   & {[}$N$, 32, 128, 128{]}           & {[}$N$, 64, 64, 64{]}              & \checkmark                                        & \checkmark       &  Leaky ReLU                    \\ \hline
$2^{\mathrm{nd}}$ contracting block   & {[}$N$, 64, 64, 64{]}             & {[}$N$, 128, 32, 32{]}             & \checkmark                                        & \checkmark          &  Leaky ReLU                     \\ \hline
$3^{\mathrm{rd}}$ contracting block   & {[}$N$, 128, 32, 32{]}            & {[}$N$, 256, 16, 16{]}             & \checkmark                                        & \checkmark                 &  Leaky ReLU              \\ \hline
$4^{\mathrm{th}}$ contracting block   & {[}$N$, 256, 16, 16{]}            & {[}$N$, 512, 8, 8{]}               & -                                        &   \checkmark       &  Leaky ReLU                         \\ \hline
$5^{\mathrm{th}}$ contracting block   & {[}$N$, 512, 8, 8{]}              & {[}$N$, 1024, 4, 4{]}               & -                                        &   \checkmark       &  Leaky ReLU              \\ \hline
$6^{\mathrm{th}}$ contracting block   & {[}$N$, 1024, 4, 4{]}             & {[}$N$, 2048, 2, 2{]}               & -                                        &   \checkmark       &  Leaky ReLU      \\ \hline
$1^{\mathrm{st}}$ expanding block     & {[}$N$, 2048, 2, 2{]}         & {[}$N$, 1024, 4, 4{]}              & -                                        &   \checkmark       &  ReLU                \\ \hline
$2^{\mathrm{nd}}$ expanding block    & {[}$N$, 1024, 4, 4{]}         & {[}$N$, 512, 8, 8{]}          & -                                        &   \checkmark       &  ReLU            \\ \hline
$3^{\mathrm{rd}}$ expanding block     & {[}$N$, 512, 8, 8{]}          & {[}$N$, 256, 16, 16{]}            & -                                        &   \checkmark       &  ReLU                 \\ \hline
$4^{\mathrm{th}}$ expanding block     & {[}$N$, 256, 16, 16{]}      & {[}$N$, 128, 32, 32{]}               & -                                        &   \checkmark       &  ReLU                    \\ \hline
$5^{\mathrm{th}}$ expanding block     & {[}$N$, 128, 32, 32{]}      & {[}$N$, 64, 64, 64{]}                & -                                        &   \checkmark       &  ReLU               \\ \hline
$6^{\mathrm{th}}$ expanding block     & {[}$N$, 64, 64, 64{]}       & {[}$N$, 32, 128, 128{]}             & -                                        &   \checkmark       &  ReLU                 \\ \hline
$2^{\mathrm{nd}}$ convolutional layer & {[}$N$, 32, 128, 128{]}           & {[}$N$, $\mathbb{C}_{out}$, 128, 128{]}             &         -                                 &          - & Linear                    \\ \hline
\end{tabular}
\end{table}

\begin{figure}
    \centering
    \includegraphics[scale=0.5]{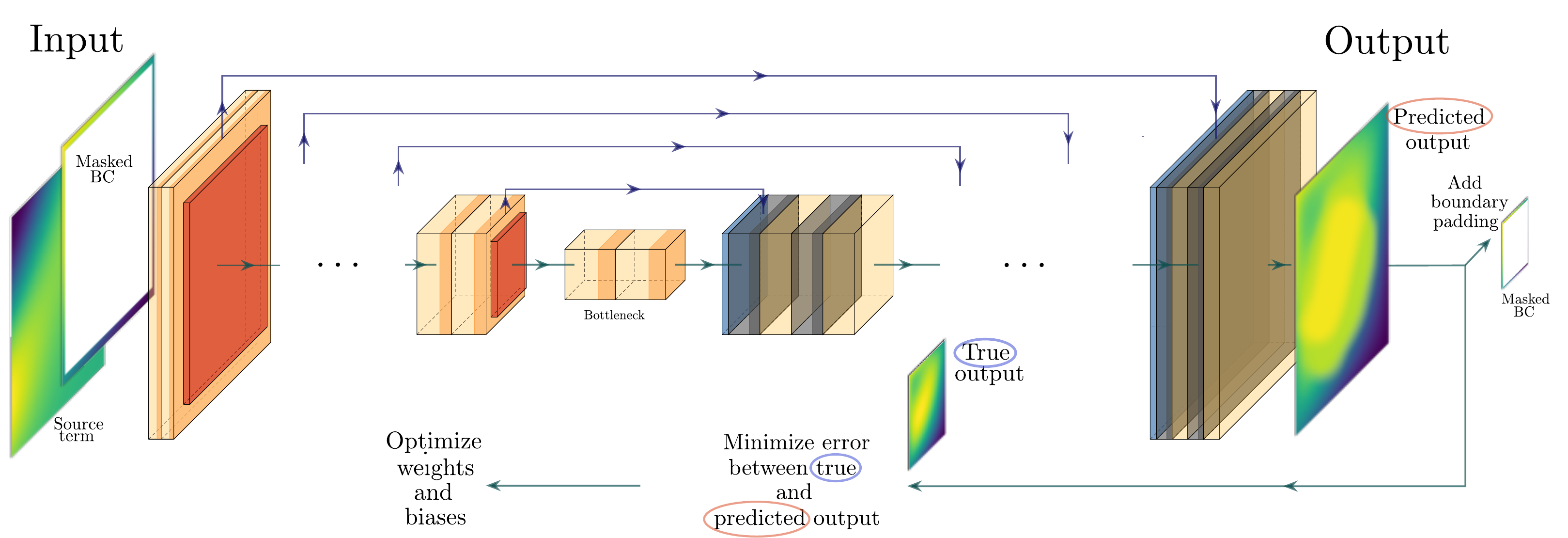}
    \caption{Schematic flow of training a CNN architecture with labeled data to build a surrogate for the Poisson equation with source term and boundary condition parametrizations. Orange and red component indicates contracting block (before bottleneck), blue and yellow component specifies expanding block (after bottleneck).}
    \label{fig:unetSchematic}
\end{figure}

Using this network and given the input and output datasets of the parameterized Poisson equation of eq. (\ref{eq::PoissonData}) we can formulate a loss function used for training the surrogate.
Since we are faced with a regression problem, we define a mean-squared error (MSE) of the form
\begin{equation}
\mathcal{L}(\bm{\theta}) =  \frac{1}{\mathbb{DOF}^{2} \, N} \sum_{i=1}^{N} \sum_{j=1}^{\mathbb{DOF}} \sum_{k=1}^{\mathbb{DOF}} \left( \hat{U}_{i,1,j,k}(\bm{\theta}) - U_{i,1,j,k} \right)^{2} + \underbrace{\text{enforcing boundary conditions}}_{\text{if not implicitly}}
\end{equation}
where $\hat{\bm{U}}$ is the output prediction of the network and $\bm{\theta}$ are the trainable parameters. The loss consists of the MSE part and a second (optional) part that is used to enforce the correct boundary conditions on the output. In this work, we implicitly enforce the boundary conditions using padding, which will be described in section \ref{sec::4}.
We can see that this is a residual error, i.e., the more accurate the surrogate model (with regards to the training data), the closer $\mathcal{L}$ gets to zero.
The optimized set of parameters of the CNN can then be obtained by solving the optimization problem
\begin{equation}\label{eq::lossCNN}
    \bm{\theta}^{\star} = \argmin_{\bm{\theta}} \mathcal{L}(\bm{\theta})
\end{equation}
which can be optimized using a stochastic gradient optimizer such as ADAM \cite{kingma2014adam}.
A flow chart for training the CNN (given labeled data of the Poisson equation)
is depicted in Figure \ref{fig:unetSchematic}.

\subsection{Convolutional physics-informed neural networks (CPINN) without labeled data}
Residual errors might be the essential component of linear and nonlinear regression analysis \cite{montgomery2021introduction,sebernonlinear90}.
Hence, it is only natural that machine learning solvers of PDEs mainly employ residual formulations, see \cite{tripathy2018deep,zhu2019physics}.
Given the output $\hat{\bm{U}}$ of a CNN and a discretized source field $\bm{F}$, the residual formulation for the Poisson equation (eq. (\ref{eq::Poisson}) reads
\begin{equation}\label{eq:PoisResidual}
 \mathcal{\bm{R}} =   \Delta \hat{\bm{U}} + \bm{F}.
\end{equation}
If the boundary conditions are fulfilled and $\text{mean}({\mathcal{\bm{R}}}) \rightarrow 0$, the output of the CNN, namely $\hat{\bm{U}}$, will be a good approximation of the true solution field $\bm{U}$. Hence, in contrast to the labeled data case of Section \ref{sec::CNNs} where the difference between the network output and the true (known) output is minimized to train the network, in CPINNs, the minimization of the residual $\mathcal{\bm{R}}$ is the goal of the training process. The requirement constrains the solution space so that the boundary conditions need to be fulfilled.

The minimization of the left-hand side of eq. (\ref{eq:PoisResidual}) requires the discretization of the Laplacian differential operator $\Delta$. In PINN formulations involving feedforward neural networks where the dependent parameters are the network's inputs, the differential operator can be resolved using automatic differentiation of the network outputs concerning the inputs. This is not an option in the current CNN formulation since the network inputs are the parameter fields of the PDEs (in the case of the Poisson equation $\bm{F}$).

We can make use of the fact that
CNN filters (c.f. Figure \ref{fig:convOperation}) are structurally equivalent to finite difference stencils \cite{mohan2020embedding}.
We can see this in the following example; consider the Laplacian operator discretized by central difference schemes where $h= 1/(\mathbb{DOF})$ represents the constant spacing variable:
\begin{equation}
\begin{aligned}
    \Delta u(x,y)&=u_{xx}(x,y)+u_{yy}(x,y)\\ & \approx{\frac {u(x-h,y)+u(x+h,y)-4u(x,y)+u(x,y-h)+u(x,y+h)}{h^{2}}}\\&=:\Delta _{h}u(x,y)
\end{aligned}
\end{equation}
which is equivalent to the commonly applied $5$-point stencil \citep{leveque2007finite}
\begin{equation}\label{eq::5PointStencil}
\begin{aligned}
        \Delta _{h}={\frac {1}{h^{2}}}{\begin{bmatrix}0 & 1 & 0\\
        1 & -4 & 1\\
        0 & 1 & 0\end{bmatrix}}.
\end{aligned}
\end{equation}
Hence, if we replace the convolutional filter (or kernel) of Figure \ref{fig:convOperation} with the stencil of eq. (\ref{eq::5PointStencil}) we can obtain a simple way to discretize the Laplacian operator spatially.
Therefore, by adding a fixed, non-trainable convolutional operation at the end of a CNN, we can approximate the residual formulation of eq. (\ref{eq:PoisResidual}).

\begin{figure}
    \centering
    \includegraphics[scale=0.5]{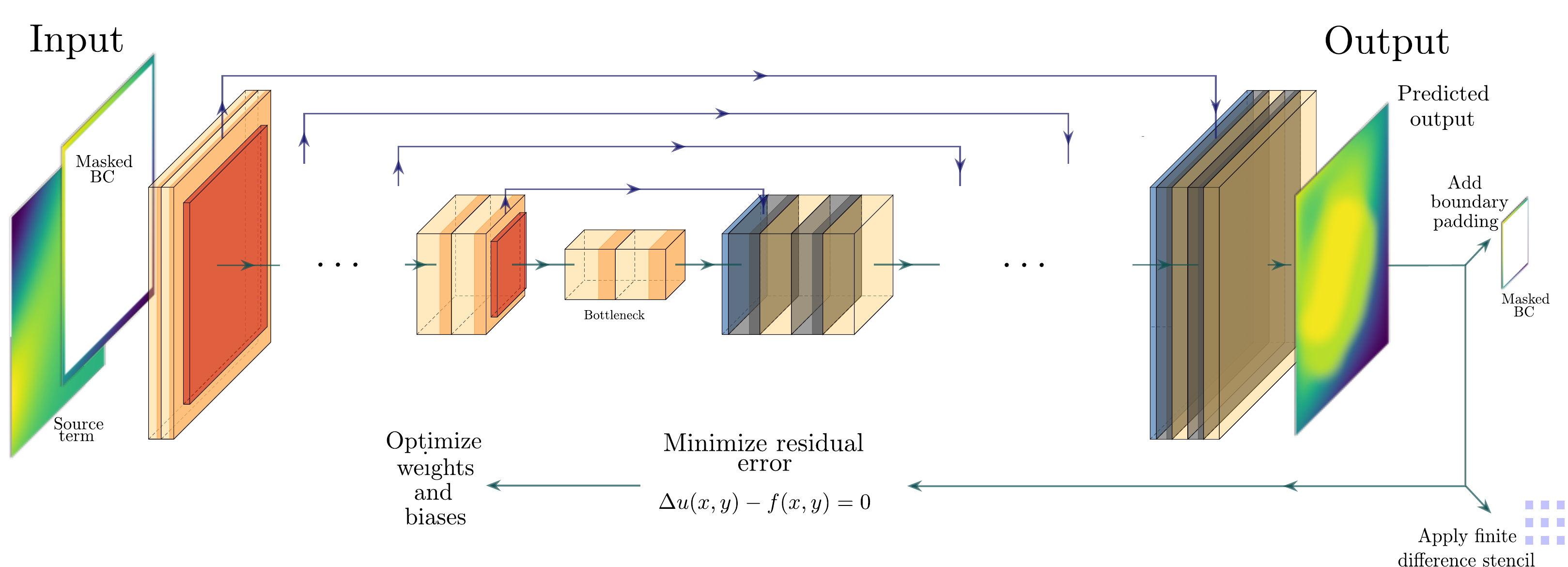}
    \caption{Schematic flow of training a CNN architecture using convolutional physics-informed neural networks (CPINN) without labeled data to build a surrogate for the Poisson equation with source term and boundary condition parametrizations. Orange and red component indicates contracting block (before bottleneck), blue and yellow component specifies expanding block (after bottleneck).}
    \label{fig:CPINNSchematic}
\end{figure}

Ultimately, the CPINN formulation can be trained by minimizing the following loss function
\begin{equation}\label{eq::lossCPINN}
\mathcal{L} =  \frac{1}{\mathbb{DOF}^{2}  \, N} \sum_{i=1}^{N} \sum_{j=1}^{\mathbb{DOF}} \sum_{k=1}^{\mathbb{DOF}} \left(  \hat{\mathcal{R}}_{i,1,j,k} \right)^{2} + \underbrace{\text{enforcing boundary conditions}}_{\text{if not implicitly}}
\end{equation}
which consists of a term for the mean squared of the approximated residual and a term that constraints the solution space by enforcing the boundary conditions.
Since we opt to implicitly enforce the boundary condition (see Section \ref{sec::4}) the latter term is automatically zero.
The CPINN approach is depicted in Figure \ref{fig:CPINNSchematic}.

\section{Deep convolutional Ritz method (DCRM) without labeled data}\label{sec::3}
So far, we have reviewed two different ways of generating surrogates for parametric PDEs (i) the classical approach, which uses a CNN trained on labeled data, and (ii) CPINN, which does not need labeled data and instead solves the PDE based on the PDE residual error. 
Here we propose an additional framework called the Deep convolutional Ritz method (DCRM).
It works essentially similar to CPINN, but
instead of finding a solution by minimizing the residual of the PDE, akin to CPINN, DCRM aims to minimize the energy functional (c.f. eq. (\ref{eq::GeneralEnergy})) which is a non-residual value. This idea has recently been explored in context of solving non-parametric PDEs  \cite{yu2018deep,samaniego2020energy,fuhg2022mixed,hamel2022calibrating}.

Energy-based solvers for PDEs were generally first proposed over 100 years ago by Walter Ritz \cite{ritz1909neue} \footnote{There is an ongoing debate whether Ritz or Rayleigh was first to conceive the idea \cite{leissa2005historical}. We do not intend the name "Deep Convolutional Ritz Method" to be controversial. Our approach is named in the style of an earlier paper \cite{yu2018deep} termed "Deep Ritz Method."  }.
From this point on, different solution techniques for PDEs (the Finite-Element Method being the most prominent) have been developed incorporating these initial ideas. Instead of requiring that the PDE holds at every material point, like in the strong form, functionals state that the conditions must only be met in an average sense. This leads to lower order differential operators in the energy functional compared to their counterparts in the strong form, but requires the evaluation of integrals that span lines, area, or volume. 

The DCRM is explained here for the particular case of the Poisson equation; however, it can similarly be applied to other energy functionals to build a solver for PDEs.
Given the CNN output $\hat{\bm{U}}$ and the source field $\bm{F}$ we can rewrite eq. (\ref{eq::EnergyPoisson}) in a tensorial form that reads
\begin{equation}\label{eq::EnergyPoissonDCRM}
    E[u] =\int_{0}^{1} \int_{0}^{1} ( \underbrace{\frac{1}{2} \norm{\nabla \hat{\bm{U}}}^{2} - \hat{\bm{U}} \bm{F} }_{I})   - \int_{\partial \Omega_{N}} \underbrace{\hat{\bm{U}}|_{\partial \Omega_{N}} \bm{g}_{N}}_{e}
\end{equation}
where $\hat{\bm{U}}|_{\partial \Omega_{N}}$ are the values of $\hat{\bm{U}}$ at the spatial positions where the Neumann boundary conditions are prescribed.
In order to minimize eq. (\ref{eq::EnergyPoissonDCRM}) an efficient way to differentiate and integrate in this framework must be established. Similarly to CPINN, convolutional filters can be used to discretize the differential operations. With $h = 1/\mathbb{DOF}$, both components of $\nabla$ can individually be discretized using central differences which read and result in
\begin{equation}
\begin{aligned}
    \frac{\partial u }{\partial x} &\approx \frac{u_{i+1,j} - u_{i-1,j}}{2 h}, \,\, &&\text{with the filter  } \,\, \frac{1}{2 h} \begin{bmatrix}
        0 & 0 & 0 \\
    -1 & 0 & 1 \\
    0 & 0 & 0
    \end{bmatrix}, \\
    \frac{\partial u }{\partial y} &\approx \frac{u_{i,j+1} - u_{i,j-1}}{2 h}, \,\, &&\text{with the filter  } \,\, \frac{1}{2 h} \begin{bmatrix}
        0 & -1 & 0 \\
    0 & 0 & 0 \\
    0 & 1 & 0
    \end{bmatrix}.
    \end{aligned}
\end{equation}
Hence, after the CNN output is obtained $\hat{\bm{U}}$, two fixed, non-trainable convolutional operations can independently be applied to approximate the  gradient operator. 
Since the CNN output is given in the form of a regular grid, the integrals of eq. (\ref{eq::EnergyPoissonDCRM}) can simply be approximated using classical numerical integration techniques yielding
\begin{equation}\label{eq::integrated}
    \hat{E} =  \sum_{i=1}^{N} \sum_{j=1}^{\mathbb{DOF}} \sum_{k=1}^{\mathbb{DOF}}  W_{j,k} I_{i,j,k}  + \sum_{i=1}^{N} \sum_{j=1}^{\mathbb{n_{BCN}}} w_{j} e_{i,j}.
\end{equation}
Here, $n_{BCN}$ denotes the degrees of freedom affected by the Neumann boundary,$W$ and $w$ are integration weights and $I$ and $e$ are defined in eq. (\ref{eq::EnergyPoissonDCRM}).
An overview of existing numerical integration techniques is given in \cite{davis2007methods}.
Without loss of generality, we employ Simpson's rule in this work. The integration weights for this method are given in
Appendix \ref{sec:AppendixNumInt}.

\begin{figure}
    \centering
    \includegraphics[scale=0.5]{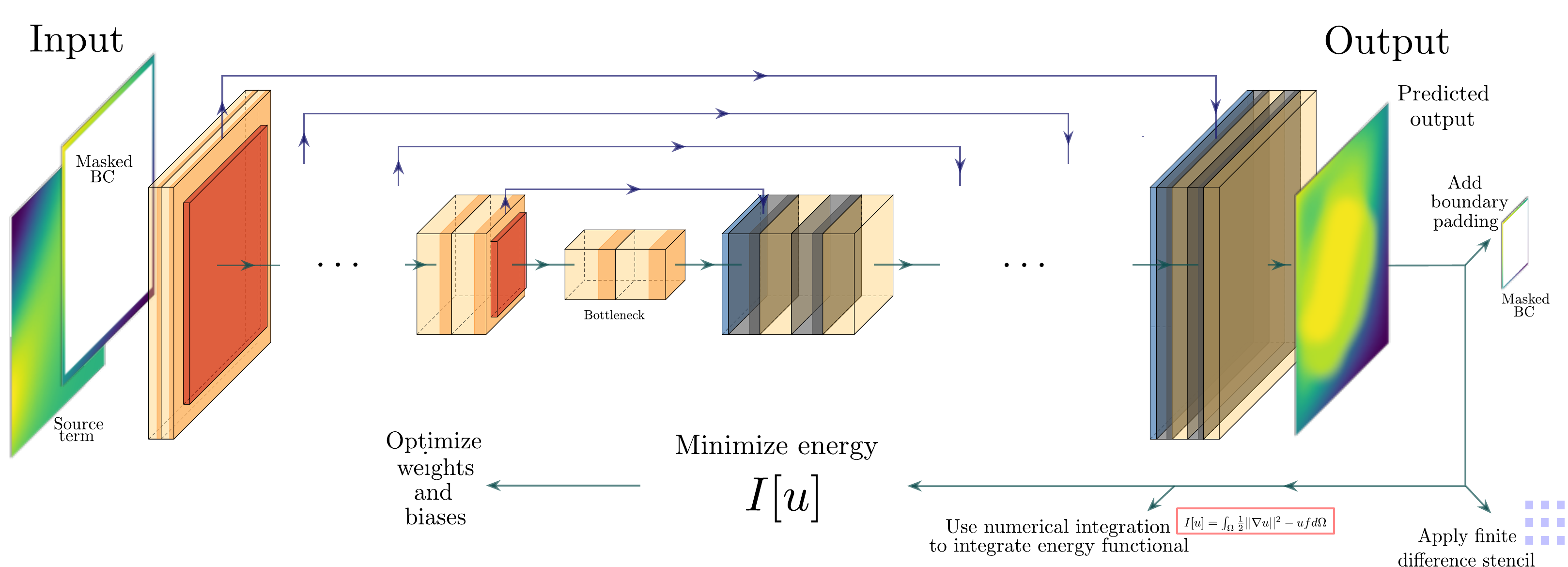}
    \caption{Schematic flow of training a Unet using the Deep convolutional Ritz method (DCRM) without labeled data to build a surrogate for the Poisson equation with source term and boundary condition parametrizations. Orange and red component indicates contracting block (before bottleneck), blue and yellow component specifies expanding block (after bottleneck).}
    \label{fig:DCRMSchematic}
\end{figure}

After obtaining an approximation $\hat{E}$ of the energy $E$ inside the system, the trainable parameters of the DCRM framework for the Poisson equation can be obtained by minimizing the loss function
\begin{equation}\label{eq::LossDCRM}
\mathcal{L} =  \hat{E}  +  \underbrace{\text{enforcing boundary conditions}}_{\text{if not implicitly}}
\end{equation}
where the combination of $\hat{E}$ and a boundary term must be minimized. A schematic overview over DCRM is provided in Figure \ref{fig:DCRMSchematic}.
One major issue with DCRM in comparison to CPINN is the fulfillment of the boundary conditions. Whereas in purely residual-based approaches, as in CNN with labeled data or CPINN, the boundary conditions could potentially be enforced in a weak sense by a penalty term to the loss function; this is not immediately possible with DCRM. This is because non-residual values ($\hat{E}$) and residual-based values (penalization of boundary conditions) are mixed in one loss function. This is problematic and would seriously dampen the potential of DCRM because the penalty parameter would have to be chosen very carefully such that the penalty term is always dominant compared to $\hat{E}$ which would lead to long training times and hyperparameter searches. To circumvent these problems, the boundary conditions will be hard encoded into the output of the CNN using a padding approach that will be explained in Section \ref{sec::4}.

\section{Hard-enforcement of boundary conditions}\label{sec::4}

In all of the discussed schemes throughout this work, the loss function (eqs. (\ref{eq::lossCNN}, \ref{eq::lossCPINN} and \ref{eq::LossDCRM})) used to train the models involved constraining the output of the CNN to adhere to the boundary conditions imposed on the PDE.
In this Section, we describe and discuss a way to enforce different boundary conditions using an alternate approach, i.e. strictly enforcing them. This contrasts with the weak imposition of the conditions using penalty approaches added to the loss.
As mentioned in the last Section, this is especially important for DCRM since non-residual and residual-based components would be combined in the loss function, which could potentially lead to problems for the optimization.
Even though enforcement of the boundary condition is not a requirement in the case where labeled data is available (Section \ref{sec::CNNs}), it is still beneficial in terms of output accuracy when specific parts of the output do not need to be learned but are implicitly enforced on the output.
Different ways of imposing boundary conditions in CNN-based surrogate modeling for PDEs are discussed in \cite{alguacil2021effects}.
In this work, we briefly discuss how to enforce Dirichlet, Neumann and periodic boundary conditions having $3\times 3$ convolutional filters in mind.

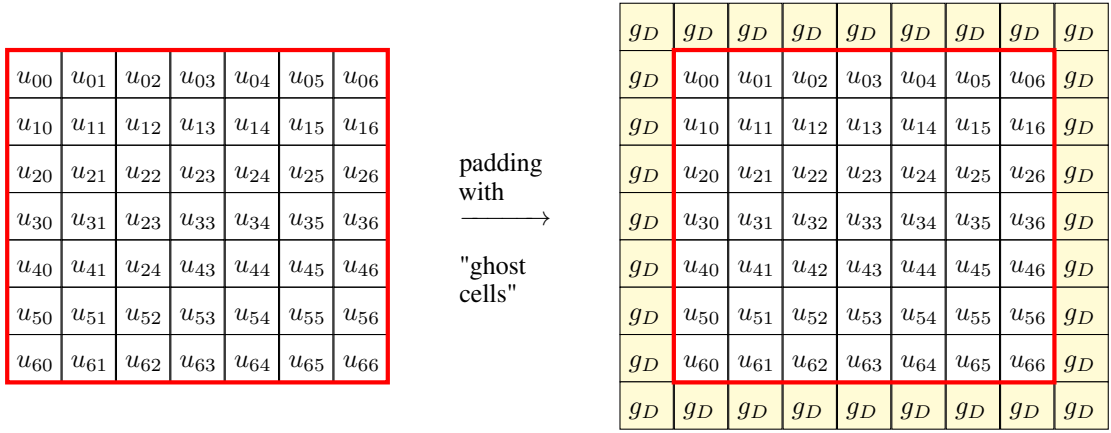
\begin{figure}
\centering
\begin{tikzpicture}[every node/.style={anchor=base,text depth=0.5ex,text height=2ex,text width=3ex}]

	\matrix (ret) [matrix of nodes,ampersand replacement=\&, row sep=-\pgflinewidth, nodes={draw}]
	{
	 	 $u_{00}$ \& $u_{01}$ \& $u_{02}$ \& $u_{03}$\& $u_{04}$ \& $u_{05}$ \& $u_{06}$\\
		$u_{10}$ \& $u_{11}$ \& $u_{12}$ \& $u_{13}$ \& $u_{14}$ \& $u_{15}$ \& $u_{16}$\\
		$u_{20}$ \& $u_{21}$ \& $u_{22}$ \& $u_{23}$ \& $u_{24}$\& $u_{25}$ \& $u_{26}$\\
		$u_{30}$ \& $u_{31}$ \& $u_{23}$ \& $u_{33}$ \& $u_{34}$ \& $u_{35}$ \& $u_{36}$ \\
		$u_{40}$ \& $u_{41}$ \& $u_{24}$ \& $u_{43}$ \& $u_{44}$ \& $u_{45}$ \& $u_{46}$ \\
		$u_{50}$ \& $u_{51}$ \& $u_{52}$ \& $u_{53}$ \& $u_{54}$ \& $u_{55}$ \& $u_{56}$ \\
		$u_{60}$ \& $u_{61}$ \& $u_{62}$ \& $u_{63}$ \& $u_{64}$ \& $u_{65}$ \& $u_{66}$\\
	};

		\node [right = 2.0em of ret] (arr1) {$\xrightarrow{\hspace*{1cm}}$};

			\node [above = 0.2em of arr1] (lim) {padding with};
			\node [below = 0.00em of arr1] (lim) {"ghost cells"};

		\draw[ultra thick, red] (ret-1-1.north west) rectangle (ret-7-7.south east);

		\matrix (ret2) [right = 4.0em of arr1,matrix of nodes,ampersand replacement=\&, row sep=-\pgflinewidth, nodes={draw}]
	{
	 	|[fill=yellow!20]|$g_{D}$ \& |[fill=yellow!20]|$g_{D}$ \& |[fill=yellow!20]|$g_{D}$ \& |[fill=yellow!20]|$g_{D}$ \& |[fill=yellow!20]|$g_{D}$ \& |[fill=yellow!20]|$g_{D}$ \& |[fill=yellow!20]|$g_{D}$ \& |[fill=yellow!20]|$g_{D}$ \& |[fill=yellow!20]|$g_{D}$\\
	 	|[fill=yellow!20]|$g_{D}$ \& $u_{00}$ \& $u_{01}$ \& $u_{02}$ \& $u_{03}$ \& $u_{04}$ \& $u_{05}$ \& $u_{06}$ \& |[fill=yellow!20]| $g_{D}$\\
		|[fill=yellow!20]|$g_{D}$ \& $u_{10}$ \& $u_{11}$ \& $u_{12}$ \& $u_{13}$ \& $u_{14}$ \& $u_{15}$ \& $u_{16}$ \& |[fill=yellow!20]|$g_{D}$\\
		|[fill=yellow!20]|$g_{D}$ \& $u_{20}$ \& $u_{21}$ \& $u_{22}$ \& $u_{23}$ \& $u_{24}$ \& $u_{25}$ \& $u_{26}$ \& |[fill=yellow!20]|$g_{D}$\\
		|[fill=yellow!20]|$g_{D}$ \&$u_{30}$ \& $u_{31}$ \& $u_{32}$ \& $u_{33}$ \& $u_{34}$ \& $u_{35}$ \& $u_{36}$ \& |[fill=yellow!20]|$g_{D}$\\
		|[fill=yellow!20]|$g_{D}$ \& $u_{40}$ \& $u_{41}$ \& $u_{42}$ \& $u_{43}$ \& $u_{44}$ \& $u_{45}$ \& $u_{46}$ \& |[fill=yellow!20]|$g_{D}$\\
		|[fill=yellow!20]|$g_{D}$ \& $u_{50}$ \& $u_{51}$ \& $u_{52}$ \& $u_{53}$ \& $u_{54}$ \& $u_{55}$ \& $u_{56}$ \& |[fill=yellow!20]|$g_{D}$\\
		|[fill=yellow!20]|$g_{D}$ \& $u_{60}$ \& $u_{61}$ \& $u_{62}$ \& $u_{63}$ \& $u_{64}$ \& $u_{65}$ \& $u_{66}$ \& |[fill=yellow!20]| $g_{D}$\\
		|[fill=yellow!20]|$g_{D}$ \& |[fill=yellow!20]|$g_{D}$ \& |[fill=yellow!20]|$g_{D}$ \& |[fill=yellow!20]|$g_{D}$ \& |[fill=yellow!20]|$g_{D}$ \& |[fill=yellow!20]|$g_{D}$ \& |[fill=yellow!20]|$g_{D}$ \& |[fill=yellow!20]|$g_{D}$ \& |[fill=yellow!20]|$g_{D}$\\
	};

			\draw[ultra thick, red] (ret2-2-2.north west) rectangle (ret2-8-8.south east);
\end{tikzpicture}
\caption{Padding assuming constant Dirichlet boundary conditions. }
\label{fig:constant boundary}
\end{figure}

\paragraph{Dirichlet boundary condition}
The exact imposition of Dirichlet boundary conditions can be achieved by adding a layer around the output of the image containing the boundary conditions that have to be enforced. Thereby, enlarging the the image from $\mathbb{DOF} \times \mathbb{DOF}$ to $\mathbb{DOF}+2 \times \mathbb{DOF} + 2$.
In the machine learning community adding additional pixels to the edge of the image is known as padding. An example of the padding approach can be seen in Figure \ref{fig:constant boundary}.
Here, the image on the left side represents the output of our CNN. Then, to enforce $g_{D}$ everywhere on the edge of the output, this value is padded to the initial image, creating the image on the right. 

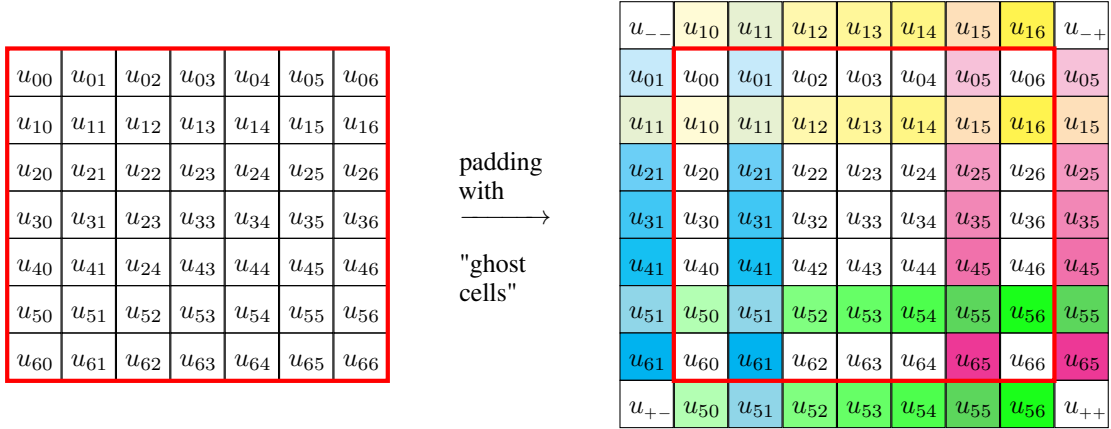
\begin{figure}
\centering
\begin{tikzpicture}[every node/.style={anchor=base,text depth=0.5ex,text height=2ex,text width=3ex}]

	\matrix (ret) [matrix of nodes,ampersand replacement=\&, row sep=-\pgflinewidth, nodes={draw}]
	{
 $u_{00}$ \& $u_{01}$ \& $u_{02}$ \& $u_{03}$\& $u_{04}$ \& $u_{05}$ \& $u_{06}$\\
		$u_{10}$ \& $u_{11}$ \& $u_{12}$ \& $u_{13}$ \& $u_{14}$ \& $u_{15}$ \& $u_{16}$\\
		$u_{20}$ \& $u_{21}$ \& $u_{22}$ \& $u_{23}$ \& $u_{24}$\& $u_{25}$ \& $u_{26}$\\
		$u_{30}$ \& $u_{31}$ \& $u_{23}$ \& $u_{33}$ \& $u_{34}$ \& $u_{35}$ \& $u_{36}$ \\
		$u_{40}$ \& $u_{41}$ \& $u_{24}$ \& $u_{43}$ \& $u_{44}$ \& $u_{45}$ \& $u_{46}$ \\
		$u_{50}$ \& $u_{51}$ \& $u_{52}$ \& $u_{53}$ \& $u_{54}$ \& $u_{55}$ \& $u_{56}$ \\
		$u_{60}$ \& $u_{61}$ \& $u_{62}$ \& $u_{63}$ \& $u_{64}$ \& $u_{65}$ \& $u_{66}$\\
	};

		\node [right = 2.0em of ret] (arr1) {$\xrightarrow{\hspace*{1cm}}$};

			\node [above = 0.2em of arr1] (lim) {padding with};
			\node [below = 0.00em of arr1] (lim) {"ghost cells"};

		\draw[ultra thick, red] (ret-1-1.north west) rectangle (ret-7-7.south east);

		\matrix (ret2) [right = 4.0em of arr1,matrix of nodes,ampersand replacement=\&, row sep=-\pgflinewidth, nodes={draw}]
	{
	 	$u_{\scriptscriptstyle --}$\& |[fill=yellow!20]| $u_{10}$ \& |[fill=cyan!30!yellow!30!]| $u_{11}$ \& |[fill=yellow!40]| $u_{12}$ \& |[fill=yellow!50]| $u_{13}$ \& |[fill=yellow!60]| $u_{14}$ \& |[fill=yellow!70!magenta!40]| $u_{15}$ \&  |[fill=yellow!80]| $u_{16}$ \&  $u_{\scriptscriptstyle -+}$ \\
  |[fill=cyan!20]| $u_{01}$ \& $u_{00}$ \& |[fill=cyan!20]|$u_{01}$ \& $u_{02}$ \& $u_{03}$ \& $u_{04}$ \& |[fill=magenta!30]|$u_{05}$ \&   $u_{06}$ \& |[fill=magenta!30]| $u_{05}$ \\
 |[fill=cyan!30!yellow!30]|$u_{11}$ \& |[fill=yellow!20]|  $u_{10}$ \& |[fill=cyan!30!yellow!30]| $u_{11}$ \& |[fill=yellow!40]|$u_{12}$ \& |[fill=yellow!50]|$u_{13}$ \& |[fill=yellow!60]|$u_{14}$\& |[fill=yellow!70!magenta!40]|$u_{15}$ \& |[fill=yellow!80]| $u_{16}$ \& |[fill=yellow!70!magenta!40]|$u_{15}$\\
  |[fill=cyan!50]| $u_{21}$ \&  $u_{20}$ \&  |[fill=cyan!50]| $u_{21}$ \& $u_{22}$ \& $u_{23}$ \& $u_{24}$\& |[fill=magenta!50]| $ u_{25}$ \&  $u_{26}$\& |[fill=magenta!50]| $u_{25}$\\
   |[fill=cyan!60]| $u_{31}$ \&  $u_{30}$ \&  |[fill=cyan!60]|$u_{31}$ \& $u_{32}$ \& $u_{33}$ \& $u_{34}$\& |[fill=magenta!60]| $u_{35}$ \&  $u_{36}$\& |[fill=magenta!60]| $u_{35}$\\
  |[fill=cyan!70]| $u_{41}$ \&  $u_{40}$ \& |[fill=cyan!70]| $u_{41}$ \& $u_{42}$ \& $u_{43}$ \& $u_{44}$\& |[fill=magenta!70]|$u_{45}$ \&  $u_{46}$\& |[fill=magenta!70]| $u_{45}$\\
 |[fill=cyan!80!green!40]|$u_{51}$ \&  |[fill=green!30]| $u_{50}$ \& |[fill=cyan!80!green!40]|$u_{51}$ \& |[fill=green!50]| $u_{52}$ \& |[fill=green!60]| $u_{53}$ \& |[fill=green!70]| $u_{54}$\& |[fill=green!80!magenta!80]| $u_{55}$ \& |[fill=green!90]| $u_{56}$\& |[fill=green!80!magenta!80]|$u_{55}$\\
|[fill=cyan!90]|$u_{61}$ \& $u_{60}$ \& |[fill=cyan!90]|$u_{61}$ \& $u_{62}$ \& $u_{63}$ \& $u_{64}$ \& |[fill=magenta!90]|$u_{65}$ \&  $u_{66}$ \& |[fill=magenta!90]| $u_{65}$ \\
  $u_{\scriptscriptstyle +-}$\& |[fill=green!30]| $u_{50}$ \& |[fill=cyan!80!green!40]| $u_{51}$ \& |[fill=green!50]|$u_{52}$ \& |[fill=green!60]| $u_{53}$\& |[fill=green!70]| $u_{54}$ \& |[fill=green!80!magenta!80]| $u_{55}$ \& |[fill=green!90]| $u_{56}$\&  $u_{\scriptscriptstyle ++}$\\
	};

			\draw[ultra thick, red] (ret2-2-2.north west) rectangle (ret2-8-8.south east);
\end{tikzpicture}
\caption{Padding assuming constant Neumann boundary $\frac{\partial u}{\partial \bm{n}} = \nabla u \cdot \bm{n} = 0.0$  conditions for second order schemes. The corner values can for example be obtained as the average of neighboring pixel values.}
\label{fig:neumann boundary}
\end{figure}

\paragraph{Neumann boundary condition}
Hard-imposition of Neumann boundary conditions is more complex compared to Dirichlet boundary conditions and are dependent on the CNN filter that is used to discretize the differential operator. Under consideration of a central difference scheme, the boundary condition can be approximated using
\begin{equation}
\begin{aligned}
        \frac{\partial u}{\partial \bm{n}} = \nabla u \cdot \bm{n} &= \begin{bmatrix}
        u_{,1}\\ u_{,2}
    \end{bmatrix} \cdot \begin{bmatrix} n_{1}\\n_{2} \end{bmatrix}
=
    \begin{bmatrix}
        \frac{u_{i+1,j} - u_{i-1,j}}{2 \Delta x} \\ \frac{u_{i,j+1} - u_{i,j-1}}{2 \Delta y}
    \end{bmatrix} \cdot \begin{bmatrix} n_{1}\\n_{2} \end{bmatrix} =g_{N}.
\end{aligned}
\end{equation}
This equation can be used to define the values of extra layers (padding) that are added to the outer edges of the CNN output to enforce the boundary condition which is similar to "ghost" nodes that are typically applied in in the finite different method.
If we, for example, consider the right-hand edge of the image with
$\bm{n} = [1,0]^{T}$ the condition reads
\begin{equation}
\begin{aligned}
 \begin{bmatrix}
        u_{,1}\\ u_{,2}
    \end{bmatrix}\cdot \begin{bmatrix} 1.0\\0.0 \end{bmatrix}
        &= \frac{u_{i+1,j} - u_{i-1,j}}{ 2 h} = g_{N} \\
        u_{i+1,j} &= 2 h g_{N} + u_{i-1,j}.
    \end{aligned}
\end{equation}
Figure \ref{fig:neumann boundary} shows an example of this operation when $g_{N}=0$ everywhere on the boundary. We can see that the corner points of the padded later are not defined by this operation. In traditional finite difference schemes specific equations would be added to the equation system to allow to seamlessly solve for the required corner values that enforce the conditions.
However, anticipating some numerical error (in case $g_{N}$ is different in the neighboring positions of the corner) we can set the corner values as the averages of neighboring boundary points which are defined by the operation described above.

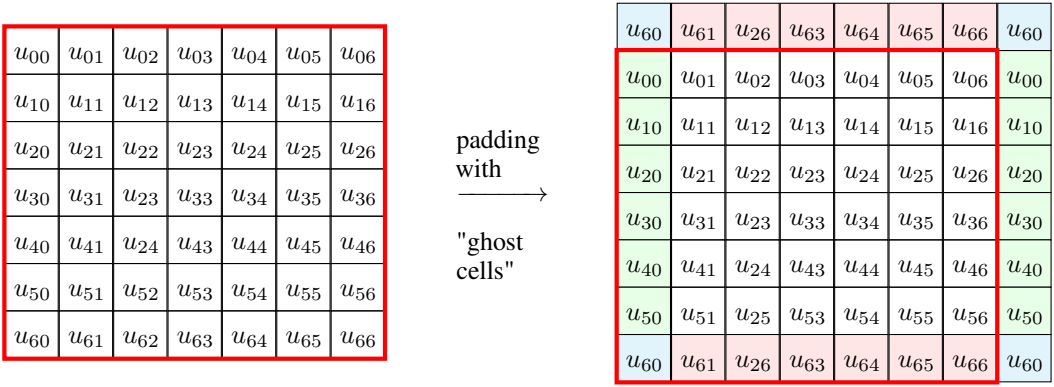
\begin{figure}
\centering
\begin{tikzpicture}[every node/.style={anchor=base,text depth=0.5ex,text height=2ex,text width=3ex}]

	\matrix (ret) [matrix of nodes,ampersand replacement=\&, row sep=-\pgflinewidth, nodes={draw}]
	{
	 	  $u_{00}$ \& $u_{01}$ \& $u_{02}$ \& $u_{03}$\& $u_{04}$ \& $u_{05}$ \& $u_{06}$\\
		$u_{10}$ \& $u_{11}$ \& $u_{12}$ \& $u_{13}$ \& $u_{14}$ \& $u_{15}$ \& $u_{16}$\\
		$u_{20}$ \& $u_{21}$ \& $u_{22}$ \& $u_{23}$ \& $u_{24}$\& $u_{25}$ \& $u_{26}$\\
		$u_{30}$ \& $u_{31}$ \& $u_{23}$ \& $u_{33}$ \& $u_{34}$ \& $u_{35}$ \& $u_{36}$ \\
		$u_{40}$ \& $u_{41}$ \& $u_{24}$ \& $u_{43}$ \& $u_{44}$ \& $u_{45}$ \& $u_{46}$ \\
		$u_{50}$ \& $u_{51}$ \& $u_{52}$ \& $u_{53}$ \& $u_{54}$ \& $u_{55}$ \& $u_{56}$ \\
		$u_{60}$ \& $u_{61}$ \& $u_{62}$ \& $u_{63}$ \& $u_{64}$ \& $u_{65}$ \& $u_{66}$\\
	};

		\node [right = 2.0em of ret] (arr1) {$\xrightarrow{\hspace*{1cm}}$};

			\node [above = 0.2em of arr1] (lim) {padding with};
			\node [below = 0.00em of arr1] (lim) {"ghost cells"};

		\draw[ultra thick, red] (ret-1-1.north west) rectangle (ret-7-7.south east);

		\matrix (ret2) [right = 4.0em of arr1,matrix of nodes,ampersand replacement=\&, row sep=-\pgflinewidth, nodes={draw}]
	{
	 	 |[fill=cyan!10]|$u_{60}$ \& |[fill=red!10]|$u_{61}$ \& |[fill=red!10]|$u_{26}$ \& |[fill=red!10]|$u_{63}$ \& |[fill=red!10]|$u_{64}$\& |[fill=red!10]|$u_{65}$ \& |[fill=red!10]|$u_{66}$ \& |[fill=cyan!10]|$u_{60}$ \\
	 |[fill=green!10]|$u_{00}$ \& $u_{01}$ \& $u_{02}$ \& $u_{03}$\& $u_{04}$\& $u_{05}$ \& $u_{06}$\& |[fill=green!10]|$u_{00}$\\
	|[fill=green!10]|$u_{10}$\& $u_{11}$ \& $u_{12}$ \& $u_{13}$ \& $u_{14}$ \& $u_{15}$ \& $u_{16}$ \& |[fill=green!10]|$u_{10}$ \\
	|[fill=green!10]|$u_{20}$ \& $u_{21}$ \& $u_{22}$\& $u_{23}$ \& $u_{24}$\& $u_{25}$ \&  $u_{26}$\& |[fill=green!10]| $u_{20}$\\
	|[fill=green!10]| $u_{30}$ \& $u_{31}$ \& $u_{23}$\& $u_{33}$ \& $u_{34}$\& $u_{35}$ \&  $u_{36}$\& |[fill=green!10]|$u_{30}$\\
	|[fill=green!10]|$u_{40}$ \& $u_{41}$ \& $u_{24}$ \& $u_{43}$ \& $u_{44}$\& $u_{45}$ \&  $u_{46}$\& |[fill=green!10]| $u_{40}$\\
	|[fill=green!10]|$u_{50}$ \& $u_{51}$ \& $u_{25}$ \& $u_{53}$ \& $u_{54}$\& $u_{55}$ \&  $u_{56}$\& |[fill=green!10]| $u_{50}$\\
	|[fill=cyan!10]| $u_{60}$ \& |[fill=red!10]| $u_{61}$ \& |[fill=red!10]| $u_{26}$ \& |[fill=red!10]| $u_{63}$ \& |[fill=red!10]| $u_{64}$\& |[fill=red!10]| $u_{65}$ \& |[fill=red!10]| $u_{66}$\& |[fill=cyan!10]| $u_{60}$ \\
	};

			\draw[ultra thick, red] (ret2-2-1.north west) rectangle (ret2-8-7.south east);
\end{tikzpicture}
\caption{Padding assuming periodic boundary conditions for second order schemes. }
\label{fig:periodic boundary}
\end{figure}

\paragraph{
Periodic boundary condition}
Periodic boundary conditions defined as
\begin{equation}
\begin{aligned}
            u_{N+1,j} &= u_{0,j} \\
            u_{i,M+1} &= u_{i,0}
\end{aligned}
\end{equation}
can be hard-encoded by adding padding layer to two sides
of the image using the values of the opposing sides.
This operation was described by \cite{mohan2020embedding}.
This approach is visually explained in Figure \ref{fig:periodic boundary} for $3\times 3$ filters.

\section{Numerical studies}\label{sec::5}
In this section, the DCRM will employ and compare CPINN and labeled-data-CNN for three different test cases involving the Poisson equation (c.f. eq. (\ref{eq::Poisson})) of increasing complexity to highlight its performance.
All network architectures were implemented in Pytorch \cite{NEURIPS20199015}. In the following, we employ the ADAM optimizer \cite{kingma2014adam} with a learning rate of $1 \times 10^{-4}$
as suggested and employed by \cite{kadeethum2021framework}.
\footnote{Codes related to this work including an implementation of DCRM will be published on Github: \url{https://github.com/FuhgJan/DCRM}, after the work has been accepted for publication.}
The labeled data for the benchmark CNN were obtained using a finite difference solver, whose solutions will also be seen as ground truths in the following. In order to allow a performance comparison of the different techniques the total absolute output error defined by
\begin{equation}
   \mathcal{E}_{abs} = \sum_{i=1}^{N} \sum_{j=1}^{\mathbb{DOF}} \sum_{k=1}^{\mathbb{DOF}} \left(  \hat{\mathcal{R}}_{i,1,j,k} \right)^{2}
\end{equation}
is used. The input fields were normalized before training.
To allow for consistent comparisons, when output data is used for training (CNN trained on labeled data), the output fields were normalized as well. The trainable parameters of all the architectures are initialized with the same values using Pytorch's default initialization.
\subsection{Case 1: Non-parametric forward solver}
The first case compares DCRM and CPINN for solving a single forward problem of the Poisson equation without any parametrization.
Consider an inhomogeneous boundary value problem of the form
\begin{equation}
\begin{aligned}
       \Delta u (x,y) &= 20 \pi^{2} (x^{2} + y^{2}) \sin (\pi (x+\frac{\pi}{4}) (y+\frac{\pi}{4})), \qquad &&\text{in}\, \Omega \\
       u &= g, \qquad &&\text{on}\, \partial\Omega \\
\end{aligned}
\end{equation}
where the Dirichlet boundary condition takes the form
\begin{equation}
    \begin{aligned}
        g &= 1, \qquad &&\text{on} \, \lbrace (x,0) \cup (x,1) \cup (0,y) \subset \partial \Omega \rbrace\\
        g &= \cos (2 \pi y), \qquad &&\text{on} \, \lbrace (1,y) \subset \partial \Omega \rbrace
    \end{aligned}
\end{equation}

\begin{figure}
    \begin{subfigure}[b]{0.5\linewidth}
    \centering
        \includegraphics[scale=0.4]{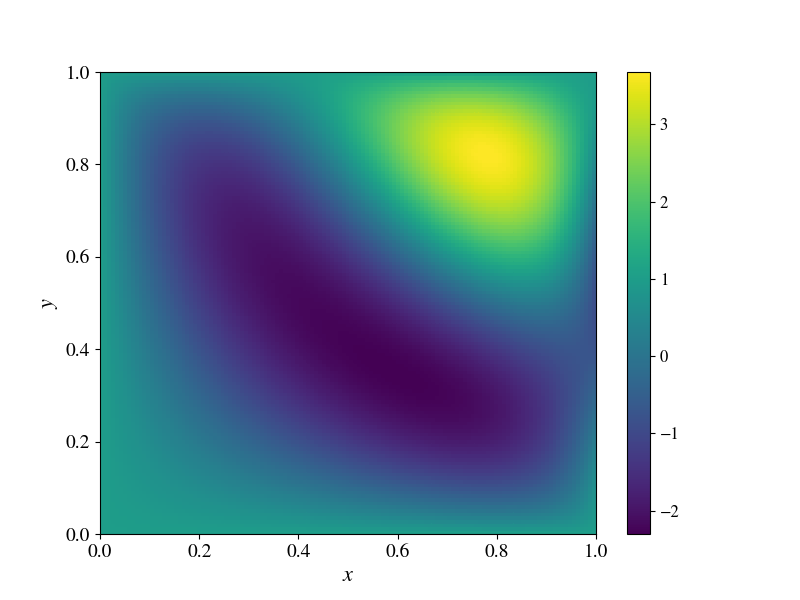}
    \caption{Ground truth}\label{fig:GroundTruthCase1}
    \end{subfigure}
    \begin{subfigure}[b]{0.5\linewidth}
        \centering
        \includegraphics[scale=0.4]{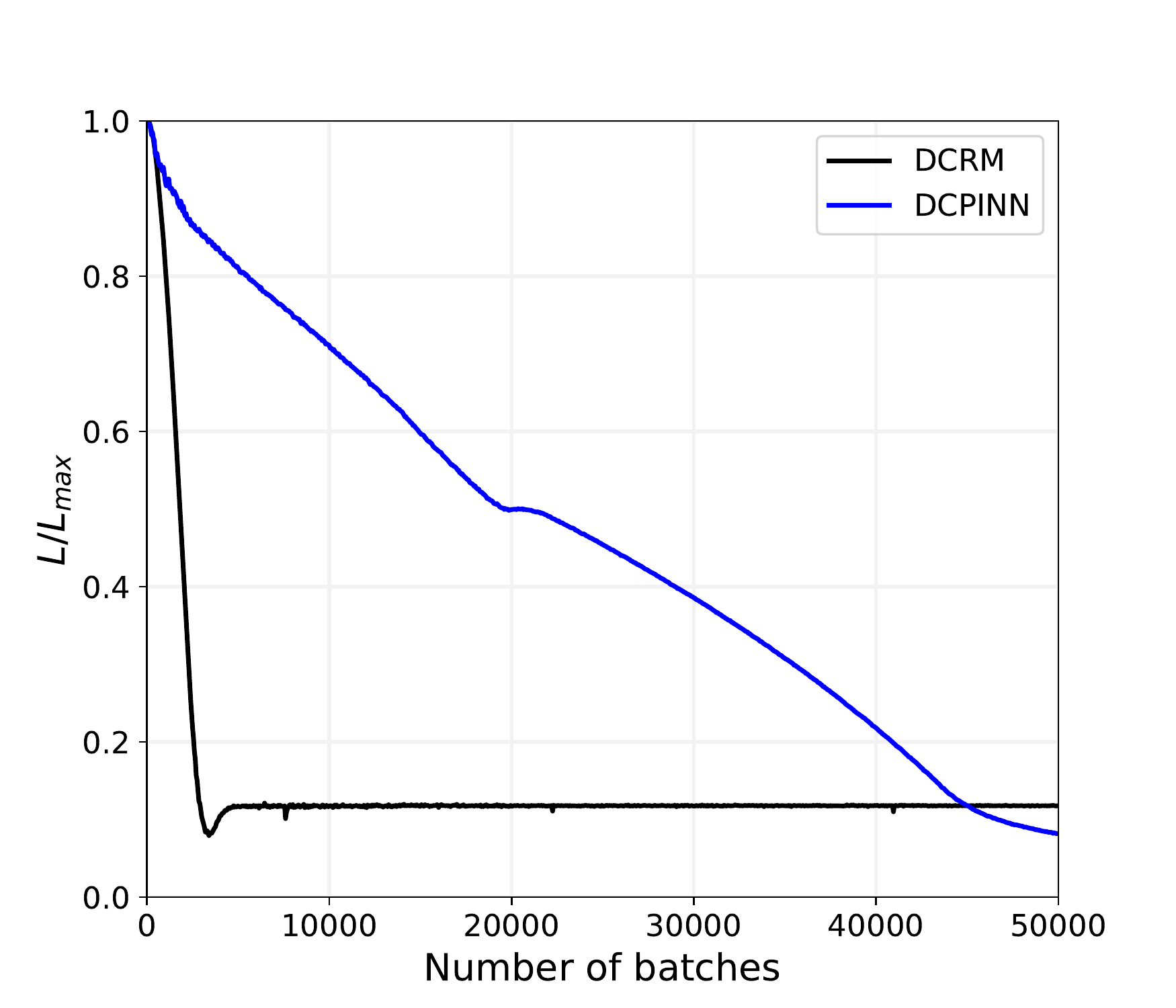}
    \caption{Training data absolute error}\label{fig:ErrorCase1}
    \end{subfigure}
    \caption{Case 1: Forward solver. (a) Ground truth solution of the forward problem, (b) Normalized absolute error between the ground truth solution and the DCRM and DCPINN predictions over the training process.}
\end{figure}

\begin{figure}
    \begin{subfigure}[b]{0.5\linewidth}
        \centering
        \includegraphics[scale=0.4]{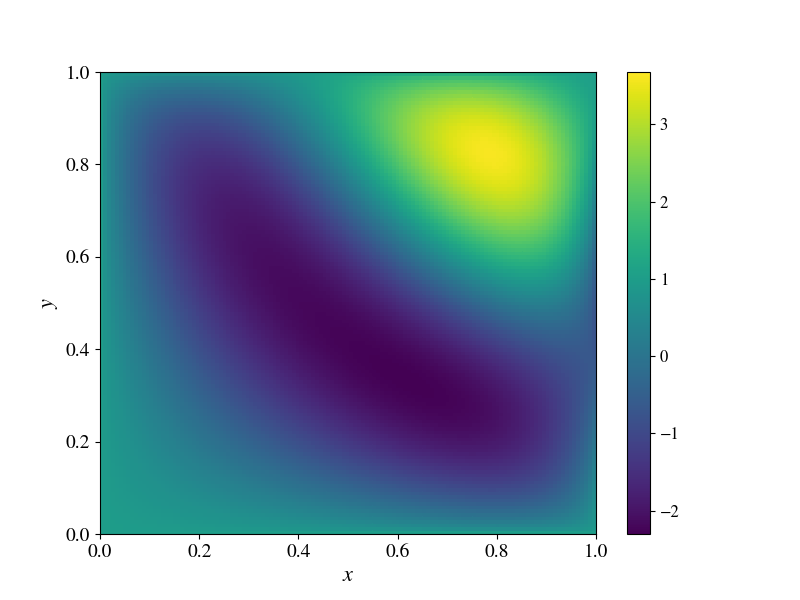}
    \caption{DCRM prediction after $3500$ epochs}\label{fig:DCRM3500}
    \end{subfigure}
    \begin{subfigure}[b]{0.5\linewidth}
        \centering
        \includegraphics[scale=0.4]{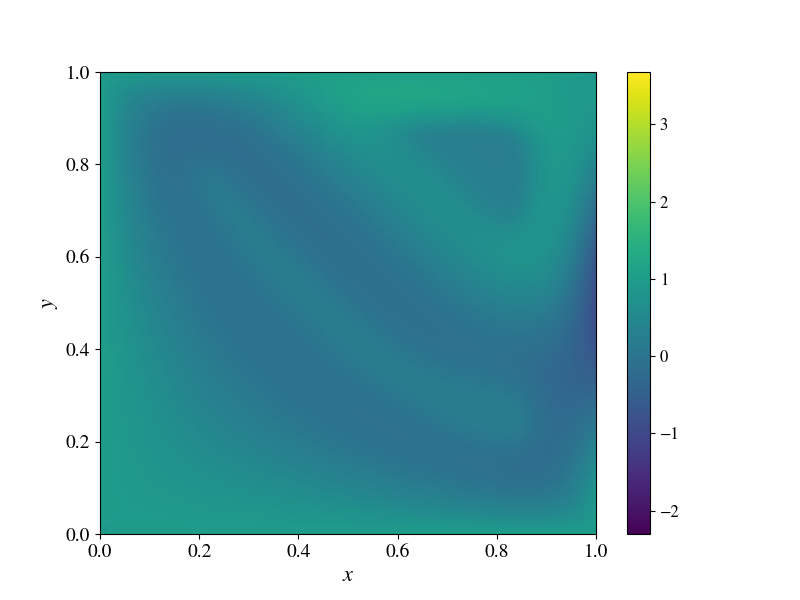}
    \caption{DCPINN prediction after $3500$ epochs}\label{fig:DCRM3500Error}
    \end{subfigure}
        \begin{subfigure}[b]{0.5\linewidth}
            \centering
        \includegraphics[scale=0.4]{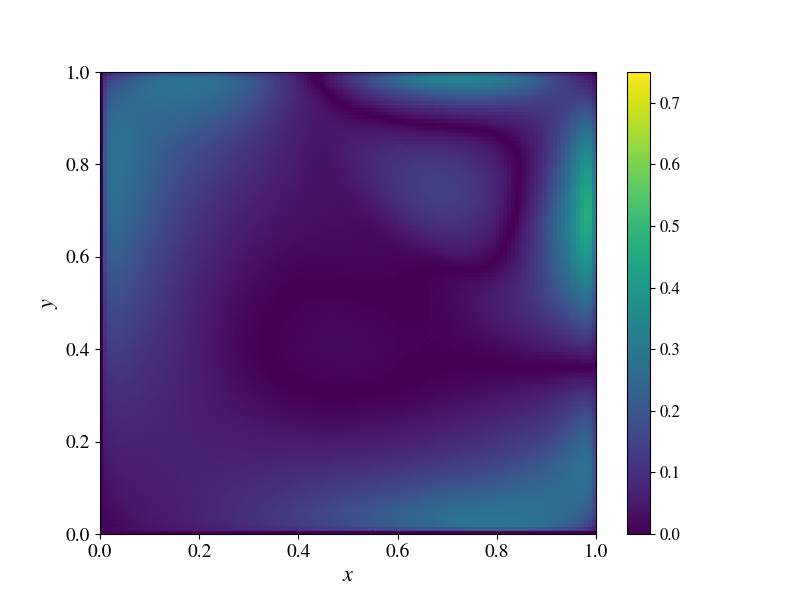}
    \caption{DCRM absolute error after $3500$ epochs}\label{fig:CPINN3500}
    \end{subfigure}
    \begin{subfigure}[b]{0.5\linewidth}
        \centering
        \includegraphics[scale=0.4]{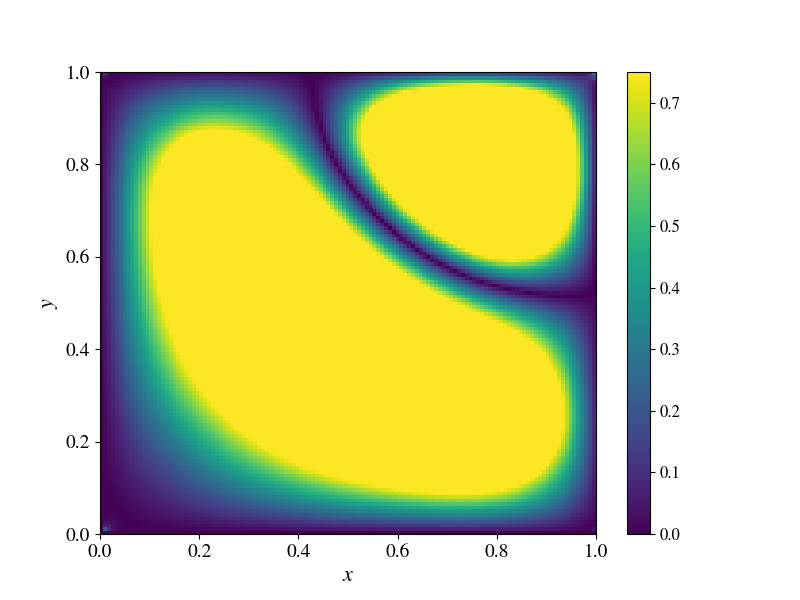}
    \caption{DCPINN absolute error after $3500$ epochs}\label{fig:CPINN3500Error}
    \end{subfigure}
    \caption{Case 1: Predicted fields after $3500$ training epochs. (a,c) DCRM predicted field and absolute error field, (b,d) DCPINN predicted field and absolute error field.}\label{fig:EvolutionCase1}
\end{figure}

The ground ground truth solution to this problem is shown in Figure \ref{fig:GroundTruthCase1}.
Figure \ref{fig:ErrorCase1} compares the absolute error evolution of DCRM and CPINN over the training process. It can be seen that DCRM initially converges to its minimum after less than $5,000$ epochs. The absolute error of CPINN eventually gets lower than DCRM but only after around $45,000$ epochs, making DCRM converge around $9$ times faster.
To put this into context
Figures \ref{fig:DCRM3500} and \ref{fig:DCRM3500Error} show the CNN output of DCRM after $3,500$ training epochs and the absolute error over the domain respectively. We can see that even after only $3,500$ epochs the output of the DCRM already closely resembles the ground truth of Figure \ref{fig:GroundTruthCase1}.
The output and absolute error of CPINN after $3,500$ epochs are depicted in Figures \ref{fig:CPINN3500} and \ref{fig:CPINN3500Error} and highlight the significant differences of the convergence speed between DCRM and CPINN.

\subsection{Case 2: Forward solver with parametrized source term}\label{sec::Case2}
\begin{figure}
    \begin{subfigure}[b]{0.5\linewidth}
        \includegraphics[scale=0.4]{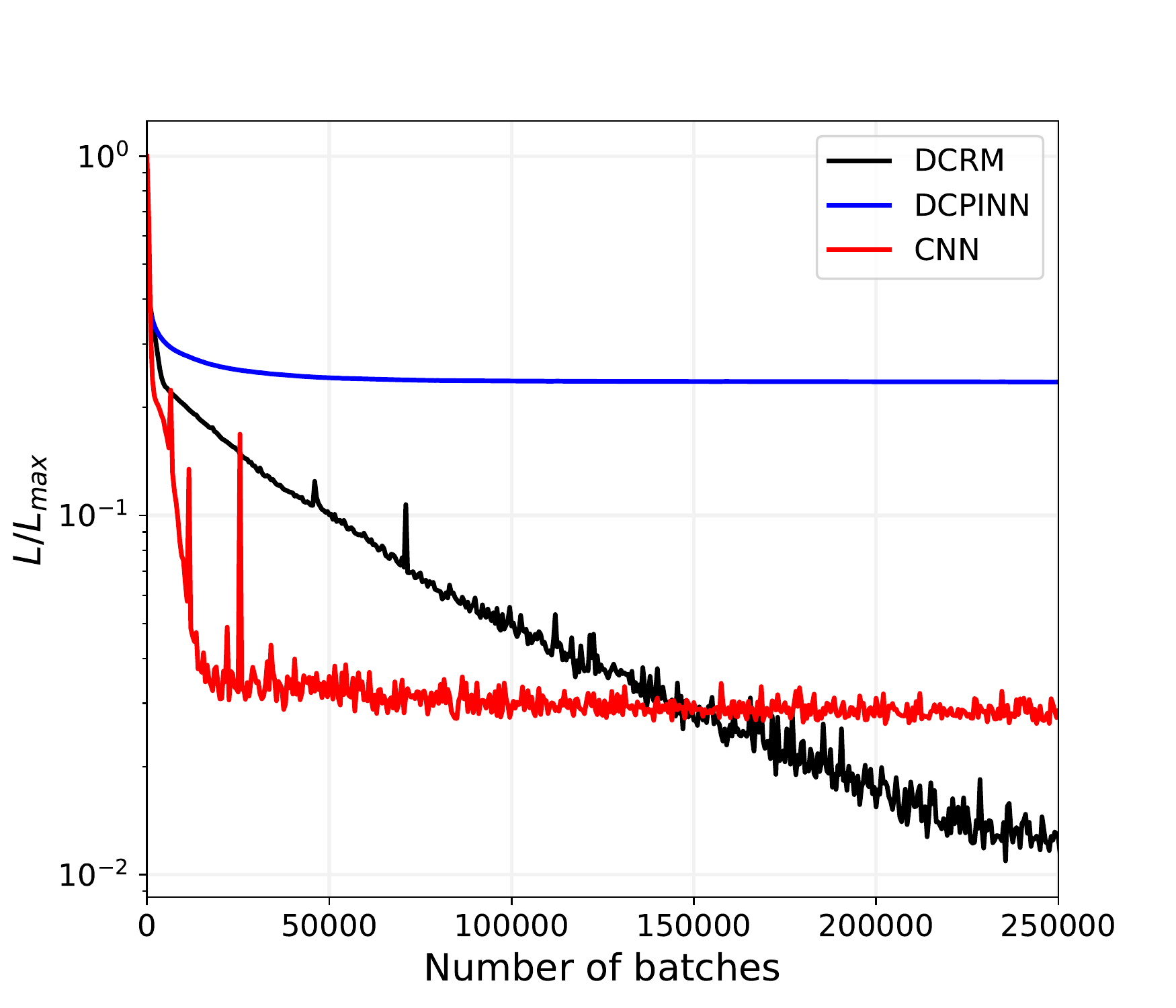}
    \caption{Normalized absolute error of training data}\label{fig:TrainErrorCase2}
    \end{subfigure}
    \begin{subfigure}[b]{0.5\linewidth}
        \includegraphics[scale=0.4]{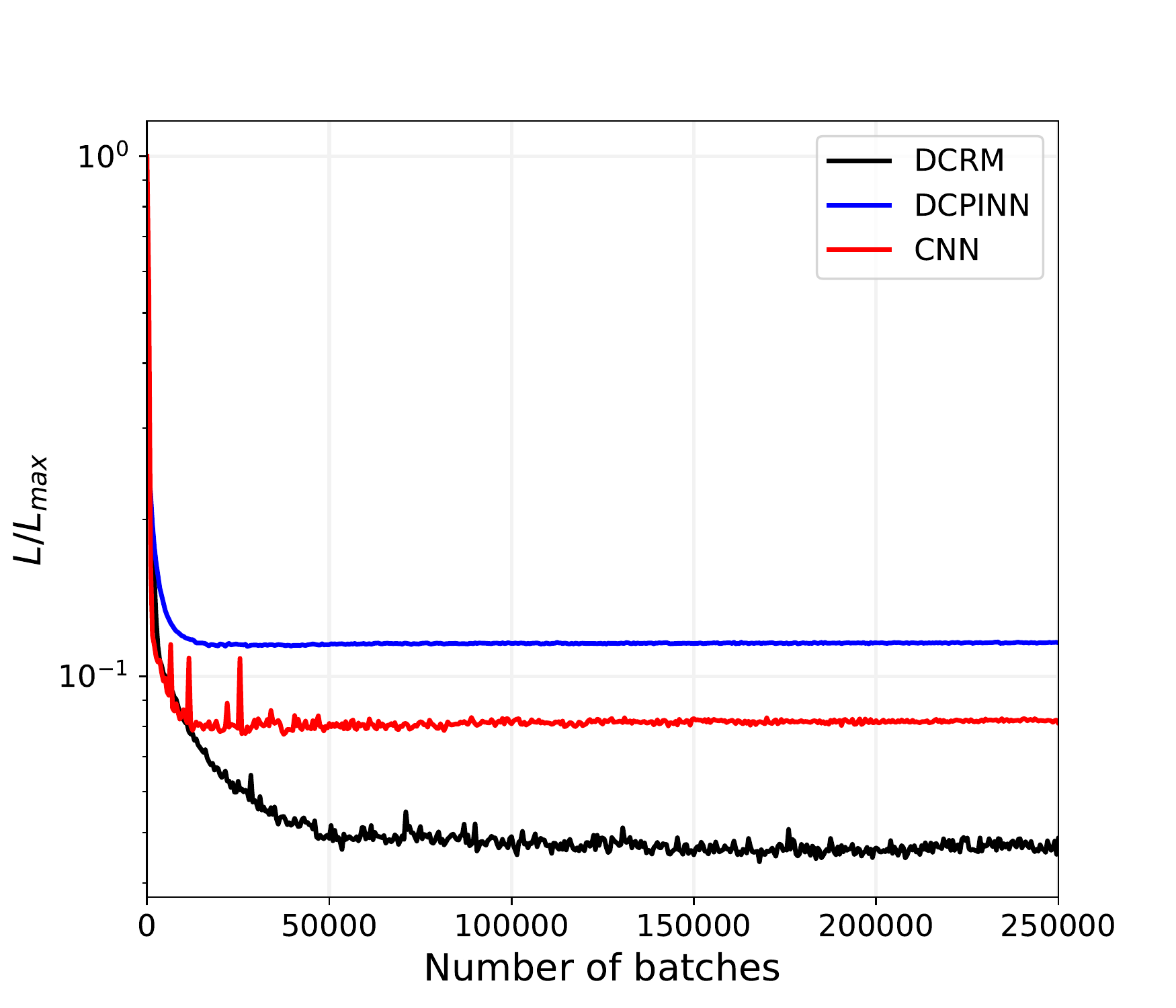}
    \caption{Normalized absolute error of test data}\label{fig:TestErrorCase2}
    \end{subfigure}
    \caption{Case 2: Transfer learning through variations in source term. Training and test errors over the training process. $100$ parametric data fields for training and $1,000$ for testing.}
\end{figure}

The second case aims to generate a surrogate solution of the PDE considering a parametrized source term of the Poisson equation. We compare the performances of DCRM to CPINN and a data-driven CNN (trained on labeled data from a FD solver).
Consider the following boundary value problem of the form
\begin{equation*}
\begin{aligned}
       \Delta u (x,y) &= f(x,y), \qquad &&\text{in}\, \Omega \\
              u &= g, \qquad &&\text{on}\, \partial\Omega \\
\end{aligned}
\end{equation*}
where the boundary conditions are specific as
\begin{equation}
    \begin{aligned}
                       g &= \cos (2 \pi y), \quad &&\text{on} \, \lbrace (1,y)  \subset  \partial \Omega \rbrace \\
        g &= 1, \quad &&\text{on} \, \lbrace \partial \Omega \setminus (1,y) \rbrace.
    \end{aligned}
\end{equation}
We obtain different source terms as samples of the following function
\begin{equation*}\label{eq::Case2SourceTerm}
    f(x,y) = \alpha \pi^{2} (x^{2} + y^{2}) \sin (\pi (x+\beta) (y+ \gamma))
\end{equation*}
where we consider
\begin{equation*}\label{eq::Case2Param}
        \alpha \in [1,10], \quad \beta \in [0, \frac{\pi}{2}], \quad \gamma \in [0, \frac{\pi}{2}]
\end{equation*}
which means that we cover a wide spectrum of complex source terms with a significant variation in magnitude and spatial variability making the process of obtaining an accurate and generalizable surrogate challenging.
We consider $100$ different samples of the source term for training purposes which are obtained using Latin Hypercube Sampling (LHS). The surrogate models are then tested on $1000$ different source terms sampled using LHS.
The models are trained using a batch size of $2$.
Figure \ref{fig:TrainErrorCase2} depicts the normalized absolute error on the training data over the training process. 

It can be seen that even though all three different approaches quickly lower the error initially, CPINN performs significantly worse than the other two methods.
CNN with labeled data initially converges the quickest, but DCRM has the lowest final error after training. The error over the test cases highlights a different performance for generalization as highlighted in
Figure \ref{fig:TestErrorCase2}. Here, even though trained without data, DCRM consistently yields the lowest error on unseen data over (essentially) the whole training process. This is a surprising result since the CNN trained on labeled data initially has a lower error on the test data. However, it seems that the physical constraints on DCRM allow it to generalize more effectively. The same can not be seen with CPINN.
Hence, at least in the studied application case, the results of CPINN compared to CNN with labeled data would suggest that the availability of (at least some) data of the input-output mapping is necessary for obtaining the best generalizing surrogate model for parametric PDEs.
However, by employing DCRM to build the surrogate model, labeled data can be made entirely obsolete with regard to generalization quality.

\subsection{Case 3:Forward solver with parametrized source term and boundary conditions}
\begin{figure}
    \centering
    \includegraphics[scale=0.9]{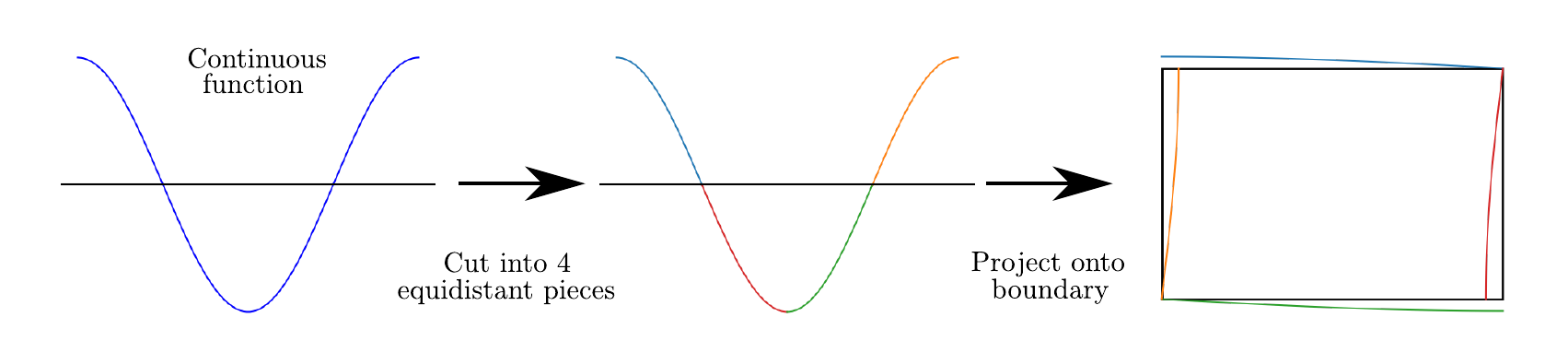}
    \caption{Parametrization of boundary condition. Projection from 1D function}
    \label{fig:parametrBoundCondition}
\end{figure}

The last case aims considers both a parameterized source term and boundary condition. We compare the DCRM to both CPINN and a CNN trained on labeled data.
The PDE in this case is given by
\begin{equation*}
\begin{aligned}
       \Delta u (x,y) &= f(x,y), \qquad &&\text{in}\, \Omega \\
               g&, \quad &&\text{on} \, \lbrace \partial \Omega \rbrace
\end{aligned}
\end{equation*}
where the Dirichlet boundary values $g$ are sampled from the following 1D function
\begin{equation*}
g(z) = \eta \cos(z), \qquad \text{with} \, z \in [0, 2\pi], \quad \text{and} \, \eta \in [-1,1]
\end{equation*}
which is projected onto the boundary of the domain, see
Figure \ref{fig:parametrBoundCondition}.
Similarly to the previous case, the source term is a realization of the function defined by
eq. (\ref{eq::Case2SourceTerm}) with the parameter values defined in the ranges of eq. (\ref{eq::Case2Param}).
The training data is generated by sampling both the boundary and source term functions $250$ times using LHS while $1,000$ test inputs are generated the same way. The models are trained using a batch-size of $2$.

\begin{figure}
    \begin{subfigure}[b]{0.5\linewidth}
        \includegraphics[scale=0.4]{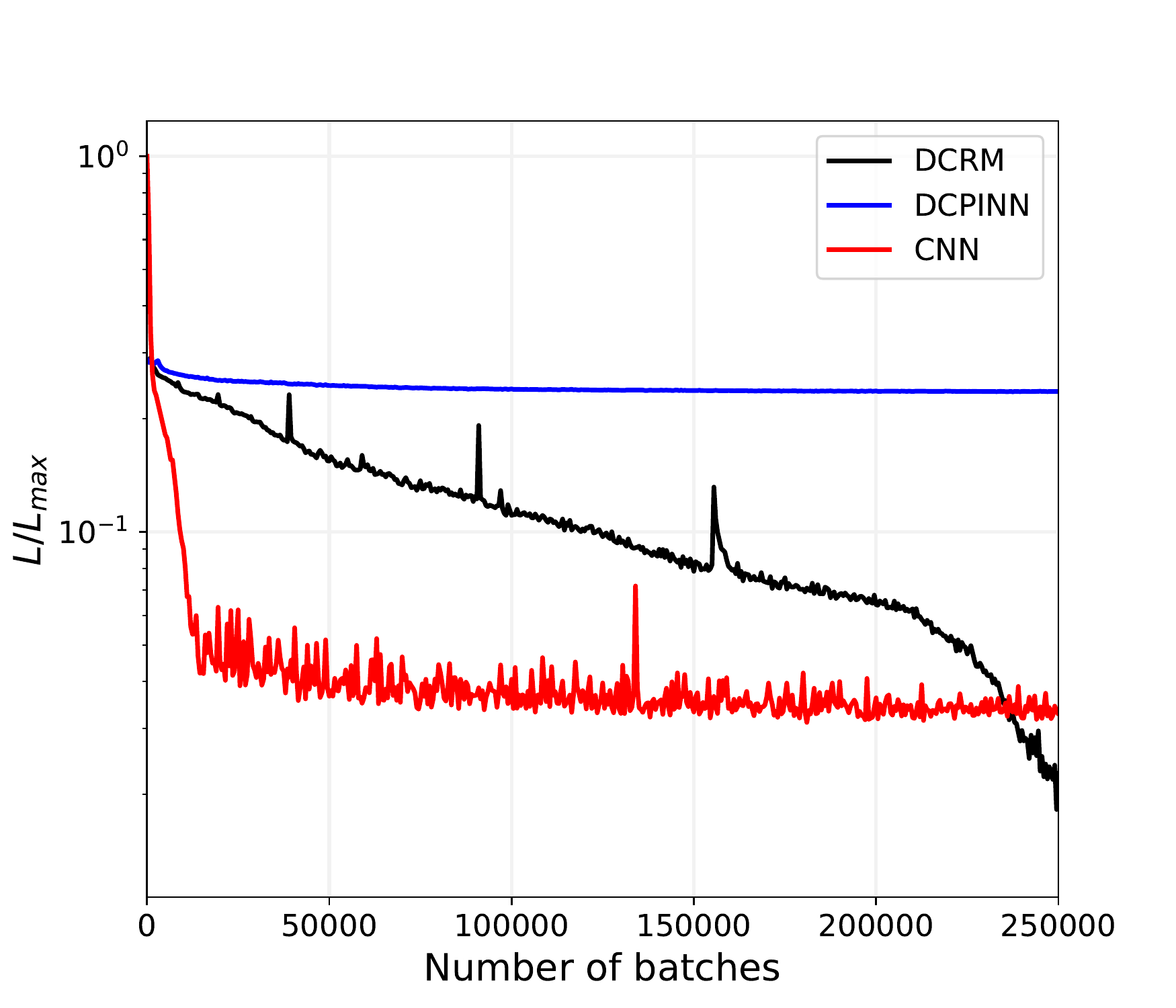}
    \caption{Absolute normalized absolute error of training data}\label{fig:TrainErrorCase3}
    \end{subfigure}
    \begin{subfigure}[b]{0.5\linewidth}
        \includegraphics[scale=0.4]{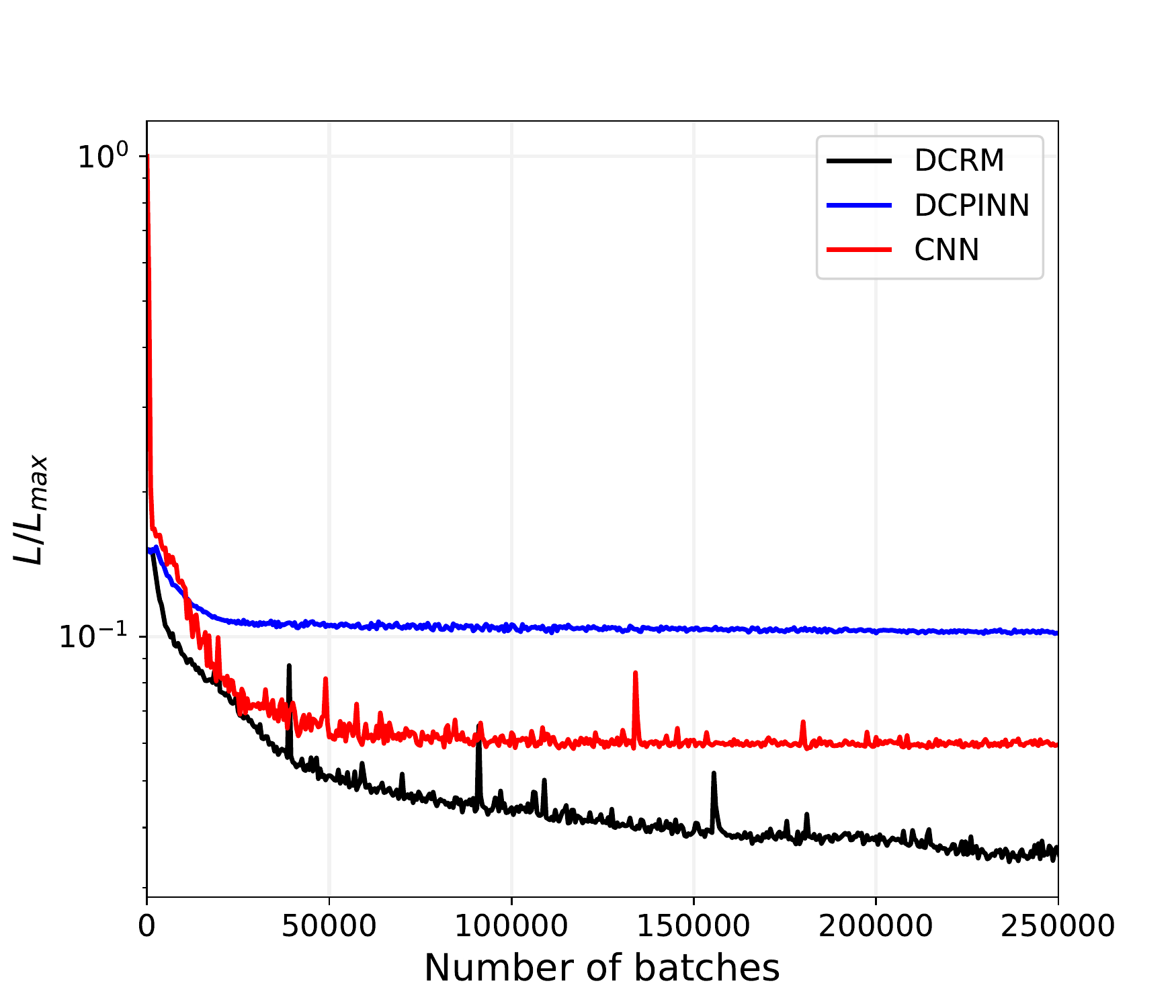}
    \caption{Absolute normalized absolute error of test data}\label{fig:TestErrorCase3}
    \end{subfigure}
    \caption{Case 3: Transfer learning through variations in source term and boundary conditions. Training and test errors over the training process. $250$ parametric data fields for training and $1,000$ for testing.}
\end{figure}

Figure \ref{fig:TrainErrorCase3} depicts the normalized error over the training data plotted over the training process. The error evolution of the three frameworks shows a similar behavior with that observed in Section \ref{sec::Case2}. The error of CPINN and CNN converge quickly, with CPINN performing worse. DCRM, on the other hand, continues to decrease until reaching a lower error level than CNN at the end of training.
The most important criterion for the quality of the surrogate models is the
generalization error. The normalized error of the test data is plotted over the training process in
Figure \ref{fig:TestErrorCase3}. We can draw the same conclusions as discussed for the previous case (Section \ref{sec::Case2}). CPINN performs worse than CNN trained with labeled data. DCRM, however, consistently has the best generalization quality over the whole training process, meaning that it performs the best on unseen data. This highlights again that (for the presented problem) DCRM can entirely cancel the need for labeled data, whereas CPINN cannot.

\section{Discussion, Conclusion and Outlook}\label{sec::6}
In this work, we propose the Deep Convolutional Ritz Method (DCRM) as a new way of generating surrogate models for parametric PDEs entirely without labeled data. The method is based on the abilities of CNNs to parametrize inputs and output fields. 

We summarize how CNNs can be used to obtain surrogate models for parametric PDEs given labeled data. Additionally, CPINNs are reviewed that allow for a generation of a surrogate model without labeled data by minimizing the residual of the PDE. In contrast to both of these methods, which rely on residual error formulations, DCRMs are trained to minimize the energy functional of the PDEs.
Here, the integrals of the functionals are approximated using numerical integration techniques, and the differential operators are resolved by adding a non-trainable, fixed convolutional filter at the output of the CNN, which is equivalent to the finite-difference operation.

The technique has a significant advantage compared to CPINNs since the degree of the differential operator is lower, making the training and accuracy of the method quicker than CPINN.
Furthermore, we discuss ways to enforce different types of boundary conditions in a strict implicit way which is  critical for DCRM since here, non-residual and residual values are combined within the loss function.
The surrogate models generated by training CNNs on labeled data, CNNs, and DCRM were tested on a Poisson equation problem with a parametric source term and boundary conditions. 
The presented results indicate that CPINNs alone do not show a better generalization ability than plain CNNs trained on labeled data as discussed in \cite{zhu2019physics}. However, DCRMs generalize better than both CPINNs and CNNs (with labeled data) without requiring any output data. Hence, using DCRM to generate a surrogate model for the parametric Poisson equation makes access to one-to-one data pairs of inputs and outputs seemingly unnecessary.

This is also emphasized by the fact that the duration of the extra steps during the training process, i.e., numerical integration and differentiation, is (time-wise) insignificant compared to the time it takes to complete a backward pass through the presented network, which is equivalent for all three compared frameworks. Hence, training a DCRM takes the same amount of time as a CNN where the labeled data is already available. However, even for this simple problem where the finite-difference solver takes seconds to evaluate, the time it takes to generate the labeled data can not be neglected., making DCRM the most efficient and best-performing way to generate a surrogate model for the parametric Poisson problem.
For boundary value problems where the numerical solver needs significant computational time, this effect might be even more pronounced. This will be the subject of future studies where surrogate models for parametric elasticity and phase-field problems
will be studied using the DCRM approach. 
Next steps could also include irregular domains and time-dependent problems, which have already been solved using regular CPINNs \cite{gao2021phygeonet}.

\section*{Acknowledgments}
This work was partly supported by the Laboratory Directed Research and Development program (218328) at Sandia National Laboratories. Sandia National Laboratories is a multimission laboratory managed and operated by National Technology \& Engineering Solutions of Sandia, LLC, a wholly owned subsidiary of Honeywell International, Inc., for the U.S. Department of Energy’s National Nuclear Security Administration under contract DE-NA0003525. This paper describes objective technical results and analysis. Any subjective views or opinions that might be expressed in the paper do not necessarily represent the views of the U.S. Department of Energy or the United States Government.

\clearpage
\bibliography{bib.bib}

\begin{thebibliography}{10}
\expandafter\ifx\csname url\endcsname\relax
  \def\url#1{\texttt{#1}}\fi
\expandafter\ifx\csname urlprefix\endcsname\relax\def\urlprefix{URL }\fi
\expandafter\ifx\csname href\endcsname\relax
  \def\href#1#2{#2} \def\path#1{#1}\fi

\bibitem{gogu2015improving}
C.~Gogu, Improving the efficiency of large scale topology optimization through
  on-the-fly reduced order model construction, International Journal for
  Numerical Methods in Engineering 101~(4) (2015) 281--304.

\bibitem{xia2014reduced}
L.~Xia, P.~Breitkopf, A reduced multiscale model for nonlinear structural
  topology optimization, Computer Methods in Applied Mechanics and Engineering
  280 (2014) 117--134.

\bibitem{keshavarzzadeh2021robust}
V.~Keshavarzzadeh, R.~M. Kirby, A.~Narayan, Robust topology optimization with
  low rank approximation using artificial neural networks, Computational
  Mechanics 68~(6) (2021) 1297--1323.

\bibitem{roache1997quantification}
P.~J. Roache, Quantification of uncertainty in computational fluid dynamics,
  Annual review of fluid Mechanics 29~(1) (1997) 123--160.

\bibitem{chen2017reduced}
P.~Chen, A.~Quarteroni, G.~Rozza, Reduced basis methods for uncertainty
  quantification, SIAM/ASA Journal on Uncertainty Quantification 5~(1) (2017)
  813--869.

\bibitem{tripathy2018deep}
R.~K. Tripathy, I.~Bilionis, Deep uq: Learning deep neural network surrogate
  models for high dimensional uncertainty quantification, Journal of
  computational physics 375 (2018) 565--588.

\bibitem{biegler2003large}
L.~T. Biegler, O.~Ghattas, M.~Heinkenschloss, B.~v. Bloemen~Waanders,
  Large-scale pde-constrained optimization: an introduction, in: Large-Scale
  PDE-Constrained Optimization, Springer, 2003, pp. 3--13.

\bibitem{fahl2003reduced}
M.~Fahl, E.~W. Sachs, Reduced order modelling approaches to pde-constrained
  optimization based on proper orthogonal decomposition, in: Large-scale
  PDE-constrained optimization, Springer, 2003, pp. 268--280.

\bibitem{zahr2015progressive}
M.~J. Zahr, C.~Farhat, Progressive construction of a parametric reduced-order
  model for pde-constrained optimization, International Journal for Numerical
  Methods in Engineering 102~(5) (2015) 1111--1135.

\bibitem{fuhg2021modeldatadriven}
J.~N. Fuhg, C.~Boehm, N.~Bouklas, A.~Fau, P.~Wriggers, M.~Marino,
  Model-data-driven constitutive responses: application to a multiscale
  computational framework (2021).
\newblock \href {http://arxiv.org/abs/2104.02650} {\path{arXiv:2104.02650}}.

\bibitem{fuhg2022local}
J.~N. Fuhg, M.~Marino, N.~Bouklas, Local approximate gaussian process
  regression for data-driven constitutive models: development and comparison
  with neural networks, Computer Methods in Applied Mechanics and Engineering
  388 (2022) 114217.

\bibitem{fuhg2022physics}
J.~N. Fuhg, N.~Bouklas, On physics-informed data-driven isotropic and
  anisotropic constitutive models through probabilistic machine learning and
  space-filling sampling, Computer Methods in Applied Mechanics and Engineering
  394 (2022) 114915.

\bibitem{wriggers2008nonlinear}
P.~Wriggers, Nonlinear finite element methods, Springer Science \& Business
  Media, 2008.

\bibitem{moukalled2016finite}
F.~Moukalled, L.~Mangani, M.~Darwish, The finite volume method, in: The finite
  volume method in computational fluid dynamics, Springer, 2016, pp. 103--135.

\bibitem{berkooz1993proper}
G.~Berkooz, P.~Holmes, J.~L. Lumley, The proper orthogonal decomposition in the
  analysis of turbulent flows, Annual review of fluid mechanics 25~(1) (1993)
  539--575.

\bibitem{couplet2005calibrated}
M.~Couplet, C.~Basdevant, P.~Sagaut, Calibrated reduced-order pod-galerkin
  system for fluid flow modelling, Journal of Computational Physics 207~(1)
  (2005) 192--220.

\bibitem{guo2018reduced}
M.~Guo, J.~S. Hesthaven, Reduced order modeling for nonlinear structural
  analysis using gaussian process regression, Computer methods in applied
  mechanics and engineering 341 (2018) 807--826.

\bibitem{ortali2020gaussian}
G.~Ortali, N.~Demo, G.~Rozza, Gaussian process approach within a data-driven
  pod framework for fluid dynamics engineering problems, arXiv preprint
  arXiv:2012.01989 (2020).

\bibitem{bhattacharya2020model}
K.~Bhattacharya, B.~Hosseini, N.~B. Kovachki, A.~M. Stuart, Model reduction and
  neural networks for parametric pdes, arXiv preprint arXiv:2005.03180 (2020).

\bibitem{zhu2018bayesian}
Y.~Zhu, N.~Zabaras, Bayesian deep convolutional encoder--decoder networks for
  surrogate modeling and uncertainty quantification, Journal of Computational
  Physics 366 (2018) 415--447.

\bibitem{kutyniok2022theoretical}
G.~Kutyniok, P.~Petersen, M.~Raslan, R.~Schneider, A theoretical analysis of
  deep neural networks and parametric pdes, Constructive Approximation 55~(1)
  (2022) 73--125.

\bibitem{khoo2021solving}
Y.~Khoo, J.~Lu, L.~Ying, Solving parametric pde problems with artificial neural
  networks, European Journal of Applied Mathematics 32~(3) (2021) 421--435.

\bibitem{sun2018discovering}
A.~Y. Sun, Discovering state-parameter mappings in subsurface models using
  generative adversarial networks, Geophysical Research Letters 45~(20) (2018)
  11--137.

\bibitem{kadeethum2021framework}
T.~Kadeethum, D.~O’Malley, J.~N. Fuhg, Y.~Choi, J.~Lee, H.~S. Viswanathan,
  N.~Bouklas, A framework for data-driven solution and parameter estimation of
  pdes using conditional generative adversarial networks, Nature Computational
  Science 1~(12) (2021) 819--829.

\bibitem{li2020neural}
Z.~Li, N.~Kovachki, K.~Azizzadenesheli, B.~Liu, K.~Bhattacharya, A.~Stuart,
  A.~Anandkumar, Neural operator: Graph kernel network for partial differential
  equations, arXiv preprint arXiv:2003.03485 (2020).

\bibitem{li2020fourier}
Z.~Li, N.~Kovachki, K.~Azizzadenesheli, B.~Liu, K.~Bhattacharya, A.~Stuart,
  A.~Anandkumar, Fourier neural operator for parametric partial differential
  equations, arXiv preprint arXiv:2010.08895 (2020).

\bibitem{liu2018survey}
H.~Liu, Y.-S. Ong, J.~Cai, A survey of adaptive sampling for global
  metamodeling in support of simulation-based complex engineering design,
  Structural and Multidisciplinary Optimization 57~(1) (2018) 393--416.

\bibitem{fuhg2021state}
J.~N. Fuhg, A.~Fau, U.~Nackenhorst, State-of-the-art and comparative review of
  adaptive sampling methods for kriging, Archives of Computational Methods in
  Engineering 28~(4) (2021) 2689--2747.

\bibitem{fuhg2022classification}
J.~N. Fuhg, A.~Fau, A classification-pursuing adaptive approach for gaussian
  process regression on unlabeled data, Mechanical Systems and Signal
  Processing 162 (2022) 107976.

\bibitem{schobi2015polynomial}
R.~Schobi, B.~Sudret, J.~Wiart, Polynomial-chaos-based kriging, International
  Journal for Uncertainty Quantification 5~(2) (2015).

\bibitem{wang2020towards}
R.~Wang, K.~Kashinath, M.~Mustafa, A.~Albert, R.~Yu, Towards physics-informed
  deep learning for turbulent flow prediction, in: Proceedings of the 26th ACM
  SIGKDD International Conference on Knowledge Discovery \& Data Mining, 2020,
  pp. 1457--1466.

\bibitem{mohan2020embedding}
A.~T. Mohan, N.~Lubbers, D.~Livescu, M.~Chertkov, Embedding hard physical
  constraints in neural network coarse-graining of 3d turbulence, arXiv
  preprint arXiv:2002.00021 (2020).

\bibitem{griewank1989automatic}
A.~Griewank, et~al., On automatic differentiation, Mathematical Programming:
  recent developments and applications 6~(6) (1989) 83--107.

\bibitem{lagaris1998artificial}
I.~E. Lagaris, A.~Likas, D.~I. Fotiadis, Artificial neural networks for solving
  ordinary and partial differential equations, IEEE transactions on neural
  networks 9~(5) (1998) 987--1000.

\bibitem{raissi2019physics}
M.~Raissi, P.~Perdikaris, G.~E. Karniadakis, Physics-informed neural networks:
  A deep learning framework for solving forward and inverse problems involving
  nonlinear partial differential equations, Journal of Computational Physics
  378 (2019) 686--707.

\bibitem{wessels2020neural}
H.~Wessels, C.~Wei{\ss}enfels, P.~Wriggers, The neural particle method--an
  updated lagrangian physics informed neural network for computational fluid
  dynamics, Computer Methods in Applied Mechanics and Engineering 368 (2020)
  113127.

\bibitem{fuhg2022interval}
J.~N. Fuhg, I.~Kalogeris, A.~Fau, N.~Bouklas, Interval and fuzzy
  physics-informed neural networks for uncertain fields, Probabilistic
  Engineering Mechanics 68 (2022) 103240.

\bibitem{lu2019deeponet}
L.~Lu, P.~Jin, G.~E. Karniadakis, Deeponet: Learning nonlinear operators for
  identifying differential equations based on the universal approximation
  theorem of operators, arXiv preprint arXiv:1910.03193 (2019).

\bibitem{lu2021learning}
L.~Lu, P.~Jin, G.~Pang, Z.~Zhang, G.~E. Karniadakis, Learning nonlinear
  operators via deeponet based on the universal approximation theorem of
  operators, Nature Machine Intelligence 3~(3) (2021) 218--229.

\bibitem{wang2021learning}
S.~Wang, H.~Wang, P.~Perdikaris, Learning the solution operator of parametric
  partial differential equations with physics-informed deeponets, Science
  advances 7~(40) (2021) eabi8605.

\bibitem{zhu2019physics}
Y.~Zhu, N.~Zabaras, P.-S. Koutsourelakis, P.~Perdikaris, Physics-constrained
  deep learning for high-dimensional surrogate modeling and uncertainty
  quantification without labeled data, Journal of Computational Physics 394
  (2019) 56--81.

\bibitem{gao2021phygeonet}
H.~Gao, L.~Sun, J.-X. Wang, Phygeonet: physics-informed geometry-adaptive
  convolutional neural networks for solving parameterized steady-state pdes on
  irregular domain, Journal of Computational Physics 428 (2021) 110079.

\bibitem{ren2022phycrnet}
P.~Ren, C.~Rao, Y.~Liu, J.-X. Wang, H.~Sun, Phycrnet: Physics-informed
  convolutional-recurrent network for solving spatiotemporal pdes, Computer
  Methods in Applied Mechanics and Engineering 389 (2022) 114399.

\bibitem{yu2018deep}
B.~Yu, et~al., The deep ritz method: a deep learning-based numerical algorithm
  for solving variational problems, Communications in Mathematics and
  Statistics 6~(1) (2018) 1--12.

\bibitem{liao2019deep}
Y.~Liao, P.~Ming, Deep nitsche method: Deep ritz method with essential boundary
  conditions, arXiv preprint arXiv:1912.01309 (2019).

\bibitem{duan2021convergence}
C.~Duan, Y.~Jiao, Y.~Lai, X.~Lu, Z.~Yang, Convergence rate analysis for deep
  ritz method, arXiv preprint arXiv:2103.13330 (2021).

\bibitem{samaniego2020energy}
E.~Samaniego, C.~Anitescu, S.~Goswami, V.~M. Nguyen-Thanh, H.~Guo, K.~Hamdia,
  X.~Zhuang, T.~Rabczuk, An energy approach to the solution of partial
  differential equations in computational mechanics via machine learning:
  Concepts, implementation and applications, Computer Methods in Applied
  Mechanics and Engineering 362 (2020) 112790.

\bibitem{fuhg2022mixed}
J.~N. Fuhg, N.~Bouklas, The mixed deep energy method for resolving
  concentration features in finite strain hyperelasticity, Journal of
  Computational Physics 451 (2022) 110839.

\bibitem{krishnapriyan2021characterizing}
A.~Krishnapriyan, A.~Gholami, S.~Zhe, R.~Kirby, M.~W. Mahoney, Characterizing
  possible failure modes in physics-informed neural networks, Advances in
  Neural Information Processing Systems 34 (2021).

\bibitem{wang2022and}
S.~Wang, X.~Yu, P.~Perdikaris, When and why pinns fail to train: A neural
  tangent kernel perspective, Journal of Computational Physics 449 (2022)
  110768.

\bibitem{anderson1980existence}
I.~M. Anderson, T.~Duchamp, On the existence of global variational principles,
  American Journal of Mathematics 102~(5) (1980) 781--868.

\bibitem{reddy2017energy}
J.~N. Reddy, Energy principles and variational methods in applied mechanics,
  John Wiley \& Sons, 2017.

\bibitem{beirao2014hitchhiker}
L.~Beir{\~a}o~da Veiga, F.~Brezzi, L.~D. Marini, A.~Russo, The hitchhiker's
  guide to the virtual element method, Mathematical models and methods in
  applied sciences 24~(08) (2014) 1541--1573.

\bibitem{douglas1941solution}
J.~Douglas, Solution of the inverse problem of the calculus of variations,
  Transactions of the American Mathematical Society 50~(1) (1941) 71--128.

\bibitem{takens1979global}
F.~Takens, A global version of the inverse problem of the calculus of
  variations, Journal of Differential Geometry 14~(4) (1979) 543--562.

\bibitem{zenkov2015inverse}
D.~Zenkov, The inverse problem of the calculus of variations, Local and Global
  Theory and Applications, Atlantis Series in Global Variational Geometry, to
  appear (2015).

\bibitem{weinstock1974calculus}
R.~Weinstock, Calculus of variations: with applications to physics and
  engineering, Courier Corporation, 1974.

\bibitem{evans1998partial}
L.~C. Evans, Partial differential equations, Graduate studies in mathematics
  19~(4) (1998) 7.

\bibitem{alguacil2021effects}
A.~Alguacil, W.~G. Pinto, M.~Bauerheim, M.~C. Jacob, S.~Moreau, Effects of
  boundary conditions in fully convolutional networks for learning
  spatio-temporal dynamics, in: Joint European Conference on Machine Learning
  and Knowledge Discovery in Databases, Springer, 2021, pp. 102--117.

\bibitem{masci2015geodesic}
J.~Masci, D.~Boscaini, M.~Bronstein, P.~Vandergheynst, Geodesic convolutional
  neural networks on riemannian manifolds, in: Proceedings of the IEEE
  international conference on computer vision workshops, 2015, pp. 37--45.

\bibitem{qi2017pointnet}
C.~R. Qi, L.~Yi, H.~Su, L.~J. Guibas, Pointnet++: Deep hierarchical feature
  learning on point sets in a metric space, Advances in neural information
  processing systems 30 (2017).

\bibitem{jiang2019convolutional}
C.~Jiang, D.~Wang, J.~Huang, P.~Marcus, M.~Nie{\ss}ner, et~al., Convolutional
  neural networks on non-uniform geometrical signals using euclidean spectral
  transformation, arXiv preprint arXiv:1901.02070 (2019).

\bibitem{gu2018recent}
J.~Gu, Z.~Wang, J.~Kuen, L.~Ma, A.~Shahroudy, B.~Shuai, T.~Liu, X.~Wang,
  G.~Wang, J.~Cai, et~al., Recent advances in convolutional neural networks,
  Pattern Recognition 77 (2018) 354--377.

\bibitem{estrach2014signal}
J.~B. Estrach, A.~Szlam, Y.~LeCun, Signal recovery from pooling
  representations, in: International conference on machine learning, PMLR,
  2014, pp. 307--315.

\bibitem{hinton2012improving}
G.~E. Hinton, N.~Srivastava, A.~Krizhevsky, I.~Sutskever, R.~R. Salakhutdinov,
  Improving neural networks by preventing co-adaptation of feature detectors,
  arXiv preprint arXiv:1207.0580 (2012).

\bibitem{srivastava2014dropout}
N.~Srivastava, G.~Hinton, A.~Krizhevsky, I.~Sutskever, R.~Salakhutdinov,
  Dropout: a simple way to prevent neural networks from overfitting, The
  journal of machine learning research 15~(1) (2014) 1929--1958.

\bibitem{ioffe2015batch}
S.~Ioffe, C.~Szegedy, Batch normalization: Accelerating deep network training
  by reducing internal covariate shift, in: International conference on machine
  learning, PMLR, 2015, pp. 448--456.

\bibitem{ronneberger2015u}
O.~Ronneberger, P.~Fischer, T.~Brox, U-net: Convolutional networks for
  biomedical image segmentation, in: International Conference on Medical image
  computing and computer-assisted intervention, Springer, 2015, pp. 234--241.

\bibitem{mao2016image}
X.~Mao, C.~Shen, Y.-B. Yang, Image restoration using very deep convolutional
  encoder-decoder networks with symmetric skip connections, Advances in neural
  information processing systems 29 (2016).

\bibitem{wang2014generalized}
W.~Wang, Y.~Huang, Y.~Wang, L.~Wang, Generalized autoencoder: A neural network
  framework for dimensionality reduction, in: Proceedings of the IEEE
  conference on computer vision and pattern recognition workshops, 2014, pp.
  490--497.

\bibitem{badrinarayanan2017segnet}
V.~Badrinarayanan, A.~Kendall, R.~Cipolla, Segnet: A deep convolutional
  encoder-decoder architecture for image segmentation, IEEE transactions on
  pattern analysis and machine intelligence 39~(12) (2017) 2481--2495.

\bibitem{hinton2006reducing}
G.~E. Hinton, R.~R. Salakhutdinov, Reducing the dimensionality of data with
  neural networks, science 313~(5786) (2006) 504--507.

\bibitem{roweis1999linear}
S.~Roweis, C.~Brody, Linear heteroencoders, Gatsby Computational Neuroscience
  Unit, Alexandra House: London, UK (1999).

\bibitem{bridgman2022heteroencoder}
W.~Bridgman, X.~Zhang, G.~Teichert, M.~Khalil, K.~Garikipati, R.~Jones, A
  heteroencoder architecture for prediction of failure locations in porous
  metals using variational inference, arXiv preprint arXiv:2202.00078 (2022).

\bibitem{kingma2014adam}
D.~P. Kingma, J.~Ba, Adam: A method for stochastic optimization, arXiv preprint
  arXiv:1412.6980 (2014).

\bibitem{montgomery2021introduction}
D.~C. Montgomery, E.~A. Peck, G.~G. Vining, Introduction to linear regression
  analysis, John Wiley \& Sons, 2021.

\bibitem{sebernonlinear90}
G.~Seber, C.~Wild, Nonlinear Regression, John Wiley \& Sons, 1990.

\bibitem{leveque2007finite}
R.~J. LeVeque, Finite difference methods for ordinary and partial differential
  equations: steady-state and time-dependent problems, SIAM, 2007.

\bibitem{hamel2022calibrating}
C.~M. Hamel, K.~N. Long, S.~L. Kramer, Calibrating constitutive models with
  full-field data via physics informed neural networks, arXiv preprint
  arXiv:2203.16577 (2022).

\bibitem{ritz1909neue}
W.~Ritz, {\"U}ber eine neue Methode zur L{\"o}sung gewisser Variationsprobleme
  der mathematischen Physik., Walter de Gruyter, Berlin/New York Berlin, New
  York, 1909.

\bibitem{leissa2005historical}
A.~W. Leissa, The historical bases of the rayleigh and ritz methods, Journal of
  Sound and Vibration 287~(4-5) (2005) 961--978.

\bibitem{davis2007methods}
P.~J. Davis, P.~Rabinowitz, Methods of numerical integration, Courier
  Corporation, 2007.

\bibitem{NEURIPS20199015}
A.~Paszke, S.~Gross, F.~Massa, A.~Lerer, J.~Bradbury, G.~Chanan, T.~Killeen,
  Z.~Lin, N.~Gimelshein, L.~Antiga, A.~Desmaison, A.~Kopf, E.~Yang, Z.~DeVito,
  M.~Raison, A.~Tejani, S.~Chilamkurthy, B.~Steiner, L.~Fang, J.~Bai,
  S.~Chintala, Pytorch: An imperative style, high-performance deep learning
  library, in: H.~Wallach, H.~Larochelle, A.~Beygelzimer, F.~d\textquotesingle
  Alch\'{e}-Buc, E.~Fox, R.~Garnett (Eds.), Advances in Neural Information
  Processing Systems 32, Curran Associates, Inc., 2019, pp. 8024--8035.

\end{thebibliography}

\clearpage
\section{Appendix}
\subsection{Numerical integration: Weights of double Trapezoid and Simpson's rule}\label{sec:AppendixNumInt}

The weights of the simple Simpson's rule are given by
\begin{equation}
    \begin{aligned}
       \bm{W} =
       \frac{1}{{3 (n-1)}}
       \begin{bmatrix}
           1 & 4& 2& 4 & \hdots & 2& 4 & 1\\
       \end{bmatrix}
    \end{aligned}.
\end{equation}
where $n_{BCN}$ is the number of degrees of interpolation points along the boundary.

The weights of the double Simpson's rule read
\begin{equation}
    \begin{aligned}
       \bm{W} =
       \frac{1}{{9 (\mathbb{DOF}-1)  (\mathbb{DOF}-1)}}
       \begin{bmatrix}
           1 & 4& 2& 4 & \hdots & 2& 4 & 1\\
           4 & 16 & 8 & 16 & \hdots & 8& 16 & 4\\
           2 & 8 & 4 & 8 & \hdots & 4 & 8 & 2 \\
           4 & 16 & 8 & 16 & \hdots & 8& 16 & 4\\
           \vdots & \vdots & \vdots & \vdots & \hdots & \vdots& \vdots & \vdots\\
                      2 & 8 & 4 & 8 & \hdots & 4 & 8 & 2 \\
                      4 & 16 & 8 & 16 & \hdots & 8& 16 & 4\\
                      1 & 4& 2& 4 & \hdots & 2& 4 & 1
       \end{bmatrix} .
    \end{aligned}
\end{equation}

\end{document}